\documentclass[aps,showpacs,preprintnumbers,amsmath, amssymb]{revtex4}

\usepackage{float}
\usepackage{graphics,epsfig}
\usepackage{graphicx}
\usepackage{dcolumn}
\usepackage{bm}
\usepackage{epstopdf}
\begin{document}
\baselineskip=0.8 cm
\title{{\bf Holographic entanglement entropy and subregion complexity for excited states of holographic superconductors }}

\author{Dong Wang$^{1}$, Xiongying Qiao$^{1}$, Mengjie Wang$^{1}$\footnote{mjwang@hunnu.edu.cn}, Qiyuan Pan$^{1,3}$\footnote{panqiyuan@hunnu.edu.cn}, Chuyu Lai$^{2}$\footnote{laichuyu@gzhu.edu.cn}, and Jiliang Jing$^{1,3}$\footnote{jljing@hunnu.edu.cn}}
\affiliation{$^{1}$Key Laboratory of Low Dimensional Quantum Structures and Quantum Control of Ministry of Education, Synergetic Innovation Center for Quantum Effects and Applications, and Department of Physics, Hunan Normal University, Changsha, Hunan
410081, China}
\affiliation{$^{2}$ Center for Astrophysics, School of Physics and Materials Science, Guangzhou University, Guangzhou 510006, China}
\affiliation{$^{3}$Center for Gravitation and Cosmology, College of Physical Science and Technology, Yangzhou University, Yangzhou 225009, China}

\vspace*{0.2cm}
\begin{abstract}
\baselineskip=0.6 cm
\begin{center}
{\bf Abstract}
\end{center}

We investigate the holographic entanglement entropy (HEE) and the holographic subregion complexity (HSC) for holographic superconductors, both in the Einstein and in the Einstein-Gauss-Bonnet gravitational theories. For both ground and excited states, we show that, in the Einstein gravity, the HSC decreases as the temperature increases and the normal phase has a smaller HSC than the superconducting phase, which is opposite to the behavior of the HEE. Moreover, we find out that, for a given temperature $T$ in the superconducting phase, the higher excited state leads to a lager value of the HEE but a smaller value of the HSC. However, the Einstein-Gauss-Bonnet gravity has significantly different effect on the HSC, while the HEE always increases monotonously as the temperature increases and its value in the normal phase always larger than that in the superconducting phase. Our results indicate that the HEE and HSC provide richer physics in phase transitions and condensation of scalar hair for holographic superconductors with excited states.

\end{abstract}

\maketitle
\newpage
\vspace*{0.2cm}

\section{Introduction}

The anti-de Sitter/conformal field theory (AdS/CFT) correspondence, more generally the gauge/gravity duality, which relates a weakly coupling gravity theory in a $(d+1)$-dimensional spacetime to a strongly coupling field theory on the $d$-dimensional boundary \cite{Maldacena,Gubser,Witten}, has been widely applied to study the strongly correlated systems in the theoretical condensed matter physics. One of its most remarkable and successful applications is providing a holographically dual description of a high temperature superconducting phase transition. Holographic superconductors can be constructed by coupling the AdS black hole with charged fields and U$(1)$ gauge fields. If the Hawking temperature is lower than some critical value, the black hole background will become unstable against perturbations and get hair by condensing some fields. According to the AdS/CFT duality, this hairy black hole solution is regarded as a superconducting phase. The first simple model of the s-wave holographic superconductor was built by Hartnoll, Herzog and Horowitz \cite{HartnollPRL,HartnollJHEP}. Considering the Yang-Mills theory/Maxwell complex vector field theory or the charged spin-two tensor field in the bulk, one can obtain the p-wave superconductors with a vector order parameter \cite{Gubser-Pufu,CaiPWave-1} and d-wave superconductors with a tensor order parameter \cite{DWaveChen,DWaveBenini} from holography. Until now, various holographic superconductor models have been constructed and have attracted tremendous interest for their potential applications to the condensed matter theory, see Refs.
\cite{CaiQongGen2015,Hartnoll2009,Herzog2009,Horowitz2011} and references therein.

Holographic superconductors with the ground state, which is the first state to condensate, have been extensively explored. While in condensed matter physics, the physical system does not necessarily remain in equilibrium, but may stay on the excited metastable states which manifest themselves in the hysteresis, superheating and supercooling phenomena, in the paramagnetic Meissner effect, in the jumps of magnetization, and in other peculiarities of the mesoscopic samples behavior, observed experimentally \cite{Zharkov}. For the mesoscopic nanomaterials, the thermal fluctuation of the system may make it turn into metastable states and the system may remain in these states for a long time because of the complicated free energy surface \cite{Peeters}. For the nanowires, the potential of superconducting nanowires lies in their long-lived excited states, which results from their low sensitivity to charge noise and critical current noise \cite{Mooij}, and the lifetime of YBa$_{2}$Cu$_{3}$O$_{7-x}$ nanowires in the excited state exceeding 20 ms at 5.4 K are superior to those in conventional Josephson junctions \cite{Lyatti}. Thus, the extension to the excited states, by considering the fact that there are many novel and important properties shown up in the excited states for superconducting materials in condensed matter systems \cite{Coffey,Sahoo,Demler,Semenov}, is interesting and significant. As the first step in this direction, Wang \emph{et al.} established a new family of solutions for holographic superconductors with excited states in the probe limit \cite{WangYQ}, and pointed out that the excited states of holographic superconductors may represent the metastable states of the mesoscopic superconductors \cite{Peeters,Vodolazov}. Subsequently, they built a fully backreaction holographic model of superconductor with excited states \cite{WangYQBackreaction}. Qiao \emph{et al.} developed a robust analytic approach to study the excited states for the holographic dual models in the AdS black hole \cite{QiaoXY2020} and soliton \cite{OuYangLiang2021} backgrounds, by including higher order terms in the expansion of the trial function. Li \emph{et al.} investigated the non-equilibrium dynamical transition process between the excited states of holographic superconductors \cite{Liran2020}. Following this line, more works on studying holographic superconductors with excited states can be found in \cite{XiangZW,PanJie,YuBao,ZhangZPJNPB,Xiang2022}.

Entanglement entropy and complexity, which are introduced from quantum information theory, play important roles in investigating quantum gravity and quantum field theory. The entanglement entropy is a powerful tool to probe phase transitions and characterize the degrees of freedom for a strongly coupled system. Holographically it can be calculated by Ryu-Takayanagi (RT) formula \cite{Ryu,Ryu2006}, which states that the entanglement entropy of CFTs is associated with the minimal area surface in the gravity side, namely
\begin{equation}\label{areaA}
\mathcal{S}=\frac{Area(\gamma_{A})}{4G_{N}}.
\end{equation}
Here $G_{N}$ denotes the Newtonian constant in the dual gravity theory, and $\gamma_{\mathcal{A}}$ is the RT minimal area surface in the bulk, which shares the same boundary $\partial_{\mathcal{A}}$ with the subregion $\mathcal{A}$. Since this dual description of the entanglement entropy has been checked for several cases, it can be applied to holographic superconductors. The initial work in this context was done by Albash and Johnson who evaluated the HEE in the s-wave holographic superconductor \cite{Albash}. Subsequently, the HEE in various superconductor models has also been studied \cite{Ogawa,XiDong,CaiRongGen027,LiLiFang,CaiRongGen088,CaiRongGen107,Kuang,Peng,YaoWeiping,Jeong2022}. The entanglement entropy turns out to be a good quantity to investigate the critical points and the order of holographic phase transitions.

However, the entanglement entropy is not enough to understand the rich geometric structures existed behind the horizon since it only grows for a very short time. Then the holographic dual of the complexity, which essentially describes the minimal number of gates of any quantum circuit to get a desired target state from a reference state, has recently been presented by Susskind \cite{Susskind2014}. The computations of the complexity in holography are refined into two concrete conjectures. The first one is the ``complexity=volume" (CV) conjecture \cite{SusskindCV2014,SusskindCV2016}, which states that the holographic complexity is in proportion to the volume of the extremal codimension-one bulk hypersurface meeting the asymptotic boundary on the desired time slice. The other one is known as ``complexity=action" (CA) conjecture \cite{BrownCAprl,BrownCAprd}. It proposes that the complexity corresponds to the on-shell bulk action in the Wheeler-DeWitt (WDW) patch which is the domain of dependence of some Cauchy surface in the bulk ending on the time slice of the boundary. In this work, we focus on the HSC proposed by Alishahiha \cite{Alishahiha}, which is another definition of holographic complexity based on the original CV conjecture. Following Alishahiha's proposal, we can evaluate the HSC by the codimension-one volume of the time-slice of the bulk geometry enclosed by the extremal codimension-two RT hypersurface used for the calculation of HEE as
\begin{equation}\label{VolumeA}
\mathcal{C}=\frac{Volume(\gamma_{A})}{8\pi LG_{N}},
\end{equation}
where $L$ represents the AdS radius. With the right choice of the length scale, the subregion complexity is able to yield significant results and avoids the complex computation since the extremal hypersurface will not touch the singularity.

Since the complexity can measure the difficulty of turning one quantum state into another, it is expected that the holographic complexity should capture the behavior of phase transitions of the boundary field theory, and there raises a great interest in investigating the complexity for different types of holographic superconductors. Many efforts have been made in employing the HSC as a probe of phase transition in the s-wave superconductor \cite{Momeni,Zangeneh,YangJNK,ChakrabortyCQG}, p-wave superconductor \cite{Fujita}, St$\ddot{u}$ckelberg superconductor \cite{Stuckelbergsuperconductor} and superconductor with nonlinear electrodynamics \cite{Shiyu2020,Lainonlinear2022}. Obviously, though the HSC is not able to probe the interior of a static black hole \cite{YuSenAn}, it still is a good parameter to characterize the superconducting phase transitions, and behaves in the different way from the entanglement entropy which means that the two quantities reflect different information of the holographic superconductor systems.

Considering that both the HEE and HSC can be used as the probes to the phase transition in holographic superconductors with the ground state, here we are aiming to investigate the HEE and HSC for excited states of the backreacting holographic superconductors in the Einstein gravity and Einstein-Gauss-Bonnet gravity. The motivation for completing this work is two folds. On one side, it is worthwhile to study the HEE and HSC for excited states of the holographic superconductors, and examine whether the HEE and HSC are still valid to reveal properties of excited states in holographic superconductor models. On the other side, both for the ground state and excited states, it would be important to unveil some general features for the HEE and HSC, and seek out good probes to the phase transition in the holographic dual models. It should be noted that the Einstein-Gauss-Bonnet gravity model is one of the natural modifications for the Einstein gravity by including Gauss-Bonnet term which arises naturally from the low-energy limit of heterotic string theory \cite{GB1,GB2,GB3,CaiRongGen2002}. Significantly, the presence of Gauss-Bonnet correction terms does not result in more than second derivatives of the metric in the corresponding field equations and thus the theory is ghost-free. The Gauss-Bonnet theory has earned much attentions in holographic studies in the past decades, and the previous works of the holographic superconductors in the Gauss-Bonnet gravity suggest that the curvature term has nontrivial contributions to some universal properties in the Einstein gravity, for example see Refs. \cite{Gregory,PanQiYuan2010,Gregory2011,Kanno,Gangopadhyay,Ghorai,Sheykhi,Salahi,LiZH,Nam,Parai,LiHF2011,LuJW,LiuSC,Mohammadi,LaiCY,NieZY}. Particularly, it was believed that the Gauss-Bonnet correction term only plays roles in spacetimes with the dimension $d\geq5$ until Glavan and Lin introduced a novel $4$-dimensional Einstein-Gauss-Bonnet gravity by rescaling the Gauss-Bonnet parameter $\alpha\rightarrow\alpha/(d-4)$ and taking the limit $d\rightarrow4$, where the curvature term makes an important contribution to the gravitational dynamics \cite{Glavan}. Subsequently, the ``regularized" versions of the $4$-dimensional Einstein-Gauss-Bonnet gravity \cite{HluPLB809,Hennigar,Fernandes,Oikonomou} and the consistent theory of $d\rightarrow4$ \cite{Aoki} have also been proposed. In Ref. \cite{QiaoXY}, the authors constructed the $(2+1)$-dimensional superconductors in the Einstein-Gauss-Bonnet gravity in the probe limit, which shows that the critical temperature $T_{c}$ first decreases then increases as the correction parameter tends towards the Chern-Simons limit in a scalar mass dependent fashion. This subtle effect of the higher curvature correction on scalar condensation for the s-wave superconductor in $(2+1)$-dimensions is quite different from the counterpart in the higher-dimensional superconductors \cite{Gregory,PanQiYuan2010,Gregory2011}. In this work, we will also study the HEE and HSC for the excited states of the superconductor models in the $4$-dimensional Gauss-Bonnet gravity away from the probe limit, which can present us some interesting details of excited states of superconductors under the impact of the Gauss-Bonnet curvature correction.

This paper is organized as follows. In section II, we investigate the HEE and HSC for excited states of holographic superconductors with fully backreaction in the Einstein gravity. In section III, we extend the discussion on the HEE and HSC of the fully backreacting holographic superconductor models to the $4$-dimensional Einstein-Gauss-Bonnet gravity. In section IV, we conclude our results.

\section{Entanglement entropy and complexity for excited states of holographic superconductors in the Einstein gravity}

\subsection{Holographic model and condensates of scalar fields}

In this subsection, we study a Maxwell field coupled with a charged complex scalar field in a $d$-dimensional Einstein gravity, via the action
\begin{eqnarray}
S=\int d^{d}x\sqrt{-g}\bigg[\frac{1}{2\kappa^{2}}(R-2\Lambda)-
\frac{1}{4}F_{\mu\nu}F^{\mu\nu}-|\nabla\psi-iqA\psi|^{2}-m^{2}|\psi|^{2}\bigg],
\end{eqnarray}
where $\kappa^{2}=8\pi G_{d}$ denotes the gravitational constant, the cosmological constant is $\Lambda=-(d-1)(d-2)/(2L^{2})$, $A$ and $\psi$ stand for the Maxwell field and a scalar with mass $m$ and charge $q$. To include the backreaction, we take the following ansatz of the metric for the black hole with a planar symmetric horizon
\begin{eqnarray}\label{metric}
ds^2&=&-f(r)e^{-\chi(r)}dt^{2}+\frac{dr^2}{f(r)}+r^{2}h_{ij}dx^{i}dx^{j}.
\end{eqnarray}
The Hawking temperature of the above black hole, which also gives the temperature of the holographic superconductor, is expressed as
\begin{eqnarray}\label{TH}
T_{H}=\frac{f^{\prime}(r_{+})e^{-\chi(r_{+})/2}}{4\pi},
\end{eqnarray}
in terms of the event horizon $r_{+}$.

Equations of motion for matter fields and metric functions, by taking $\psi=\psi(r)$ and $A=\phi(r)dt$, are given by
\begin{eqnarray}\label{chieq}
\chi^{\prime}+\frac{4\kappa^{2}r}{d-2}\bigg(\psi'^2+\frac{q^{2}e^{\chi}\phi^{2}\psi^{2}}{f^{2}}\bigg)=0,
\end{eqnarray}
\begin{eqnarray}\label{feq}
f^{\prime}-\bigg[\frac{(d-1)r}{L^{2}}-\frac{(d-3)f}{r}\bigg]+
\frac{2\kappa^{2}r}{d-2}\bigg[m^{2}\psi^{2}+\frac{e^{\chi}\phi'^2}{2}+f\bigg(\psi'^2+\frac{q^{2}e^{\chi}\phi^{2}\psi^{2}}{f^{2}}\bigg)\bigg]=0,
\end{eqnarray}
\begin{eqnarray}\label{phi}
\phi^{\prime\prime}+\bigg(\frac{d-2}{r}+\frac{\chi^{\prime}}{2}\bigg)\phi^{\prime}-\frac{2q^{2}\psi^{2}}{f}\phi=0,
\end{eqnarray}
\begin{eqnarray}\label{psi}
\psi^{\prime\prime}+\bigg(\frac{d-2}{r}+\frac{f^{\prime}}{f}-\frac{\chi^{\prime}}{2}\bigg)\psi^{\prime}+\bigg(\frac{q^{2}e^{\chi}\phi^{2}}{f^{2}}-\frac{m^{2}}{f}\bigg)\psi=0,
\end{eqnarray}
where the prime denotes the derivative with respect to the coordinate $r$. Just as in Ref. \cite{PanQY2012}, we take the unit with $q=1$ and keep $\kappa^{2}$ finite when the backreaction is taken into account.

For the normal phase, there is no condensate, i.e., $\psi(r)=0$, which leads to, from Eqs.~\eqref{chieq}--\eqref{psi}, a constant $\chi$ and an AdS Reissner-Nordstr$\ddot{o}$m (RN) black hole
\begin{eqnarray}
f(r)&=&\frac{r^{2}}{L^{2}}-\frac{1}{r^{d-3}}\bigg[\frac{r_{+}^{d-1}}{L^{2}}+\frac{(d-3)\kappa^{2}\rho^{2}}{(d-2)r_{+}^{d-3}}\bigg]+\frac{(d-3)\kappa^{2}\rho^{2}}{(d-2)r^{2(d-3)}},\nonumber \\
\phi(r)&=&\mu-\frac{\rho}{r^{d-3}}.
\end{eqnarray}
Note that here $\mu$ and $\rho$ are introduced to describe the chemical potential and the charge density in the dual field theory. In the limit of $\kappa=0$, the Schwarzschild AdS black hole is recovered.

To get the solutions of the superconducting phase, i.e., $\psi(r)\neq 0$, physically relevant boundary conditions have to be imposed. At the horizon $r=r_{+}$, the metric coefficient $\chi$ and scalar field $\psi$ are regular, but the metric coefficient $f$ and gauge field $\phi$ obey $f(r_{+})=0$ and $\phi(r_{+})=0$, respectively. Near the boundary $r\rightarrow\infty$, the asymptotic behaviors of the solutions can be expressed as
\begin{eqnarray}
\chi\rightarrow 0,~~f\sim\frac{r^{2}}{L^{2}},~~\phi\sim\mu-\frac{\rho}{r^{d-3}},
~~\psi\sim\frac{\psi_{-}}{r^{\lambda_{-}}}+\frac{\psi_{+}}{r^{\lambda_{+}}},
\end{eqnarray}
where $\psi_{+}$ and $\psi_{-}$ are related to the vacuum expectation value of the boundary operator $\mathcal{O}$ with the conformal dimension $\lambda_{\pm}=[(d-1)\pm\sqrt{(d-1)^{2}+4m^{2}L^{2}}]/2$, respectively. For the case of $\lambda_{-}$ is larger than the unitarity bound, both modes are normalizable and thus we take the boundary condition that either $\psi_{-}$ or $\psi_{+}$ vanishes \cite{HartnollPRL,HartnollJHEP}.

The following scaling symmetries and the transformations of relevant quantities
\begin{eqnarray}\label{scaling}
r\rightarrow \beta r,~~~~(t,x^{i})\rightarrow \frac{1}{\beta}(t,x^{i}),~~~~(\chi,\psi,L)\rightarrow (\chi,\psi,L),\nonumber \\
(\phi,\mu, T)\rightarrow \beta(\phi,\mu, T),~~~~\rho\rightarrow \beta^{d-2}\rho,~~~~\psi_{\pm}\rightarrow \beta^{\lambda_{\pm}}\psi_{\pm},
\end{eqnarray}
with a real positive number $\beta$, are guaranteed by Eqs.~\eqref{chieq}--\eqref{psi}. Thus, we can choose $r_{+}=1$ and $L=1$. For concreteness, we focus on the 4-dimensional AdS black hole spacetime, with $\kappa=0.05$ and $m^2L^2=-2$ satisfying the Breitenlohner-Freedman (BF) bound ($m^2L^2\geq-9/4$ for $d=4$). In numerics, we take the coordinate transformation $r\rightarrow z=r_{+}/r$ for convenience.


\begin{figure}[H]
\includegraphics[scale=0.65]{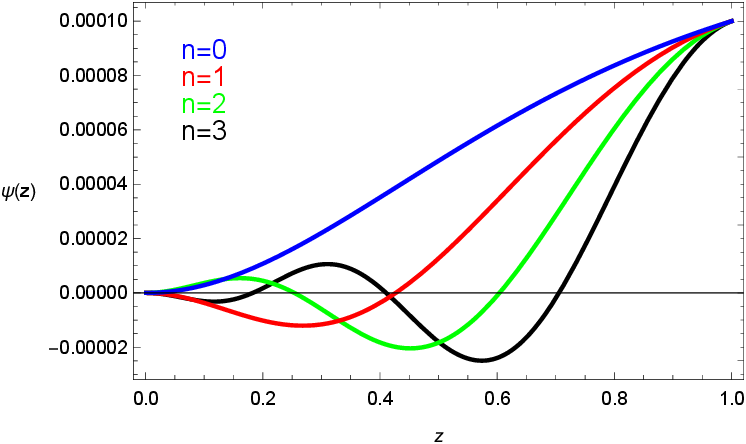}\hspace{0.2cm}%
\includegraphics[scale=0.65]{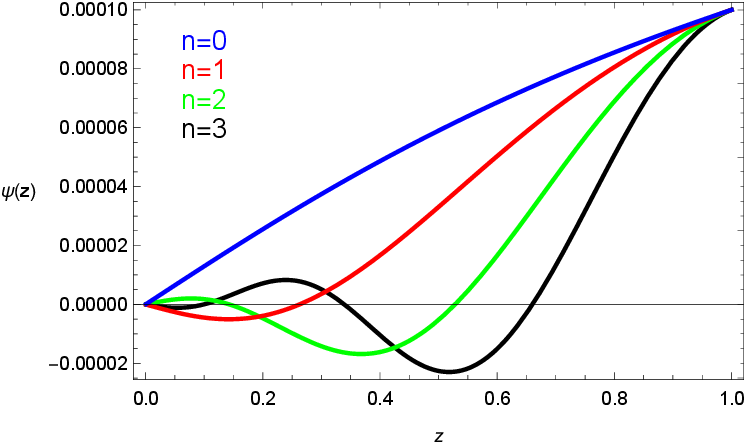}\\ \vspace{0.0cm}
\caption{\label{Psiz} (Color online) The scalar field $\psi(z)$ as a function of the radial coordinate $z$ outside the horizon with the scalar operators $\mathcal{O}_{+}$ (left) and $\mathcal{O}_{-}$ (right) for the fixed mass $m^{2}L^{2}=-2$. In each panel, the blue, red, green and black lines denote the ground $(n=0)$, first $(n=1)$, second $(n=2)$ and third $(n=3)$ states, respectively.}
\end{figure}

In Fig. \ref{Psiz}, we set the initial condition $\psi(1)=0.0001$ and plot the distribution of the scalar field $\psi(z)$ as a function of $z$ for the scalar operators $\mathcal{O}_{+}$ (left) and $\mathcal{O}_{-}$ (right) with the fixed mass of the scalar field $m^{2}L^{2}=-2$ by using the numerical shooting method. Obviously, the ``excited" states correspond to the bulk solutions for which the scalar field changes sign along the radial direction, which means that the excited states are characterized by the number of nodes of the scalar field and the ground state corresponds to the scalar field without nodes. In Ref. \cite{WangYQ}, Wang \emph{et al.} argued that these excited states of the holographic superconductors could be related to the metastable states of the mesoscopic superconductors. Thus, it is interesting and important to further consider such a configuration since we want to investigate the metastable states for superconducting materials in condensed matter systems by holography.

\begin{figure}[ht]
\includegraphics[scale=0.62]{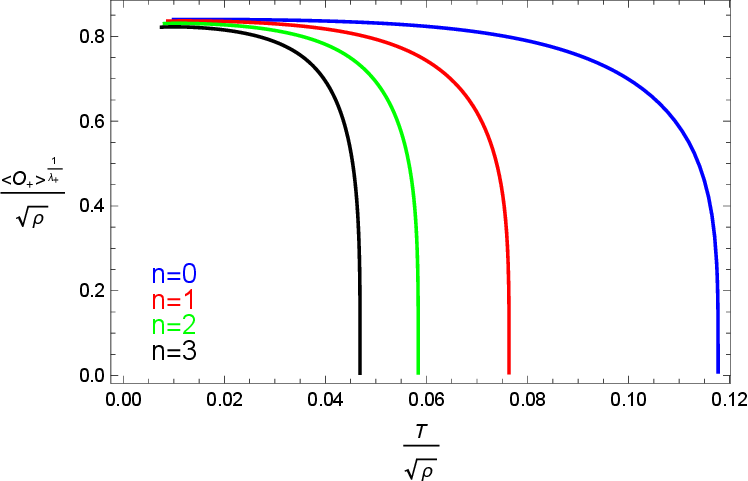}\hspace{0.4cm}%
\includegraphics[scale=0.62]{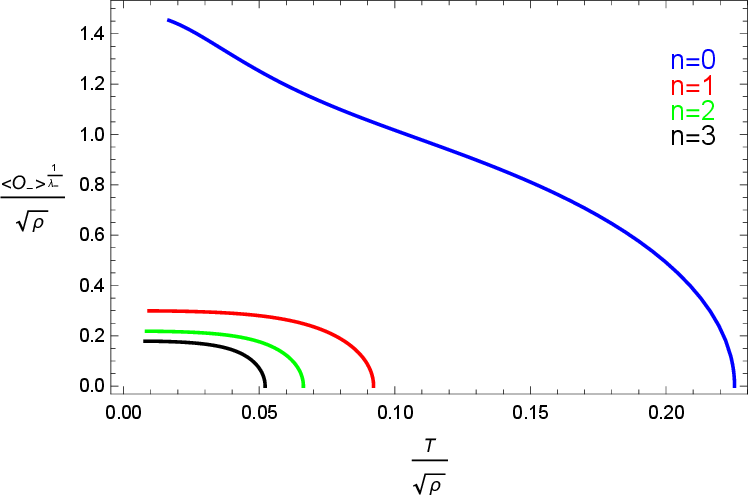}\hspace{0.4cm}%
\caption{\label{4dCond} The condensates of scalar operators $\mathcal{O}_{+}$ (left) and $\mathcal{O}_{-}$ (right) with excited states versus temperature for the fixed mass $m^{2}L^{2}=-2$. In each panel, the blue, red, green and black lines denote the ground $(n=0)$, first $(n=1)$, second $(n=2)$ and third $(n=3)$ states, respectively.}
\end{figure}

\begin{table}[ht]
\caption{The critical temperature $T_{c}$ of scalar operators $\mathcal{O}_{+}$ and $\mathcal{O}_{-}$ with excited states for the fixed mass $m^{2}L^{2}=-2$.}\label{4detemp}
\begin{center}
\begin{tabular}{c c c c c c}
         \hline
$n$ & 0 & 1 & 2 & 3 & 4
        \\
        \hline
~~~~~~$<\mathcal{O}_{+}>$~~~~~&~~~~~$0.117710\rho^{1/2}$~~~~~&~~~~~$0.076345\rho^{1/2}$~~~~~&~~~~~$0.058377\rho^{1/2}$~~~~~&~~~~~$0.046861\rho^{1/2}$~~~~~&~~~~~$0.038202\rho^{1/2}$
          \\
~~~~~~$<\mathcal{O}_{-}>$~~~~~&~~~~~$0.225271\rho^{1/2}$~~~~~&~~~~~$0.092168\rho^{1/2}$~~~~~&~~~~~$0.066290\rho^{1/2}$~~~~~&~~~~~$0.052181\rho^{1/2}$~~~~~&~~~~~$0.042283\rho^{1/2}$
          \\
        \hline
\end{tabular}
\end{center}
\end{table}

In Fig. \ref{4dCond}, we exhibit the condensates of scalar operators $\mathcal{O}_{+}$ (left) and $\mathcal{O}_{-}$ (right), from the ground state to the third excited state, versus the temperature, which shows that the condensates emerge when the temperature is lower than the critical temperature $T_c$. The critical temperatures $T_{c}$ for both operators, from the ground state to the fourth excited state, are listed in Table \ref{4detemp}. It is shown that $T_{c}$ of excited states is lower than that of the ground state, which means that the higher excited state makes it harder for the scalar hair to form. The results agree well with the findings in Ref. \cite{WangYQBackreaction}. Fitting the curves in Fig. \ref{4dCond}, we have the condensate behavior of operators as $\langle\mathcal{O}_{\pm}\rangle\sim(1-T/T_{c})^{1/2}$ near $T_{c}$, which tells us that for the ground and excited states, the superconducting phase transition of the $4$-dimensional backreacting holographic model belongs to the second order and the critical exponent is the mean field value $1/2$.

\subsection{HEE and HSC in the holographic model}

In this section, we will numerically study the behaviors of the HEE and HSC in the metal/superconductor phase transition with excited states, which will give more physics about the superconducting phase transition of excited states.

Let us consider a subsystem $\mathcal{A}$ with a straight strip geometry described by $-l/2\leq x\leq l/2$ and $-R/2\leq y\leq R/2$ $(R\rightarrow\infty)$. Here $l$ is defined as the size of region $\mathcal{A}$, and $R$ represents a regulator which will be set to infinity. For a UV cutoff $\epsilon$, the radial minimal surface $\gamma_{A}$ starts from $z=\epsilon$ at $x=l/2$, and extends into the bulk until it reaches $z=z_{\ast}$, then returns back to the AdS boundary $z=\epsilon$ at $x=-l/2$. Hence, the induced metric on the minimal surface takes the form
\begin{eqnarray}\label{inducedmetric}
ds_{induced}^{2}=\frac{r_{+}^{2}}{z^{2}}\bigg\{\bigg[1+\frac{1}{z^{2}f}\bigg(\frac{dz}{dx}\bigg)^{2}\bigg]dx^{2}+dy^{2}\bigg\}.
\end{eqnarray}
By using the RT formula given in Eq. (\ref{areaA}), we get the entanglement entropy
\begin{eqnarray}\label{RTformula}
\mathcal{S}=\frac{R}{4G_{4}}\int_{-\frac{l}{2}}^{\frac{l}{2}}\frac{r_{+}^{2}}{z^{2}}\sqrt{1+\frac{1}{z^{2}f}\bigg(\frac{dz}{dx}\bigg)^{2}}dx.
\end{eqnarray}
The minimality condition indicates
\begin{eqnarray}\label{xeq}
\frac{dz}{dx}=\frac{1}{z}\sqrt{(z_{\ast}^{4}-z^{4})f},
\end{eqnarray}
which satisfies the constraint condition $\frac{dz}{dx}|_{z=z_{\ast}}=0$ for the constant $z_{\ast}$. Setting $x(z_*)=0$, we integrate the condition (\ref{xeq}) and obtain
\begin{eqnarray}\label{RTxz}
x(z)=\int_{z}^{z_{\ast}}\frac{z}{\sqrt{(z_{\ast}^{4}-z^{4})f}}dz,
\end{eqnarray}
with $x(\epsilon\rightarrow 0)=l/2$. After minimizing the area by Eq. (\ref{xeq}), the HEE becomes
\begin{eqnarray}\label{HEE}
\mathcal{S}=\frac{R}{2G_{4}}\int_{\epsilon}^{z_{\epsilon}}\frac{z_{\ast}^{2}}{z^{3}\sqrt{(z_{\ast}^{4}-z^{4})f}}dz= \frac{R}{2G_{4}}\left(s+\frac{1}{\epsilon}\right),
\end{eqnarray}
where $s$ is the finite term and $1/\epsilon$ is the divergent term. We will subtract this divergent term from $\mathcal{S}$ in Eq. (\ref{HEE}), and analyze the physically important finite part $s$ of the HEE.

Following the proposal given by Eq. (\ref{VolumeA}), we find the HSC in the strip geometry
\begin{eqnarray}\label{HSC}
\mathcal{C}=\frac{R}{4\pi LG_{4}}\int_{\epsilon}^{z_{\ast}}\frac{x(z)dz}{z^{4}f}=\frac{R}{4\pi LG_{4}}\left[c+\frac{F(z_{\ast})}{\epsilon^{2}}\right],
\end{eqnarray}
with a universal term $c$ and a divergent term in the form of $F(z_{\ast})/\epsilon^{2}$. Note that the function $F(z_{\ast})$ has different forms under different situations so that we cannot give the general analytical expression of the HSC divergent term and subtract it off to find the universal part of $\mathcal{C}$. Fortunately, the universal term $c$ is independent of the UV cutoff. So by taking two different cutoffs $\epsilon_1$ and $\epsilon_2$, one may numerically compute the value of $F(z_{\ast})$ by
\begin{eqnarray}
F(z_{\ast})=\frac{4\pi LG_{4}[\mathcal{C}(\epsilon_{1})-\mathcal{C}(\epsilon_{2})]}{R(\epsilon_{1}^{-2}-\epsilon_{2}^{-2})},
\end{eqnarray}
which can help us to pick up the universal term $c$ of the HSC in every situation.

\begin{figure}[ht]
\includegraphics[scale=0.62]{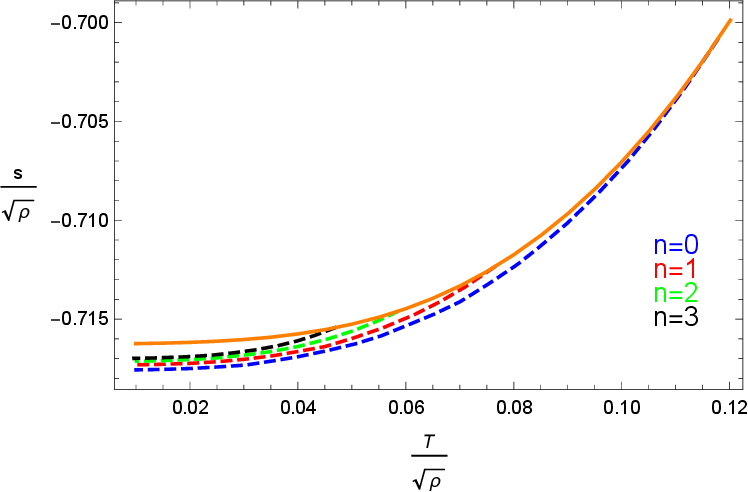}\hspace{0.4cm}%
\includegraphics[scale=0.62]{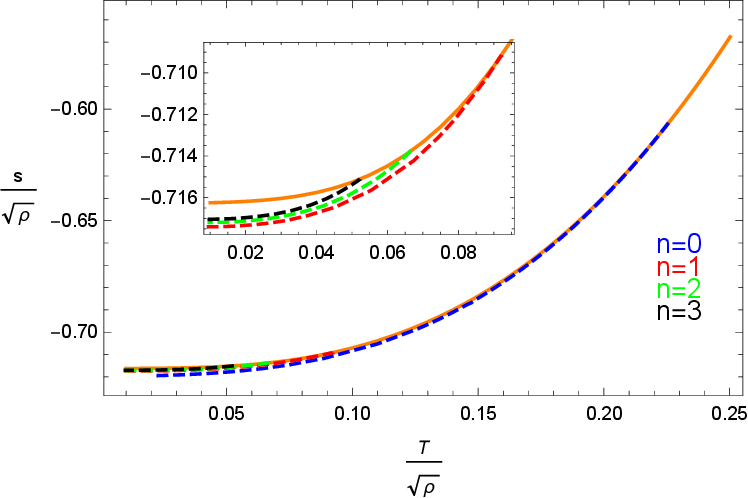}\hspace{0.4cm}%
\includegraphics[scale=0.62]{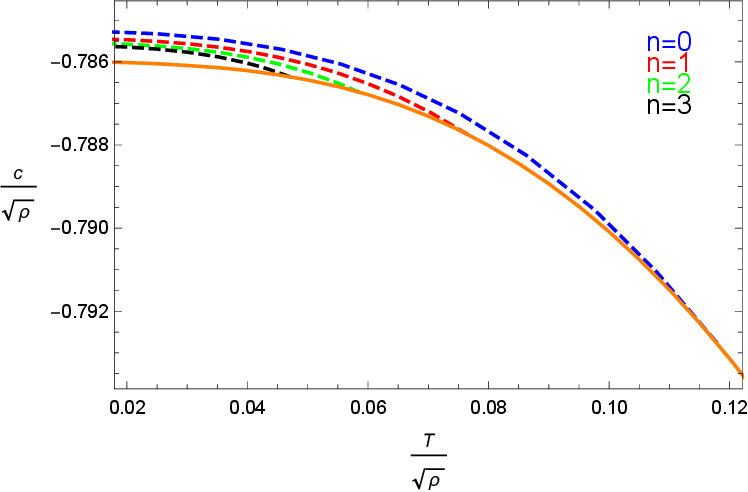}\hspace{0.40cm}%
\includegraphics[scale=0.62]{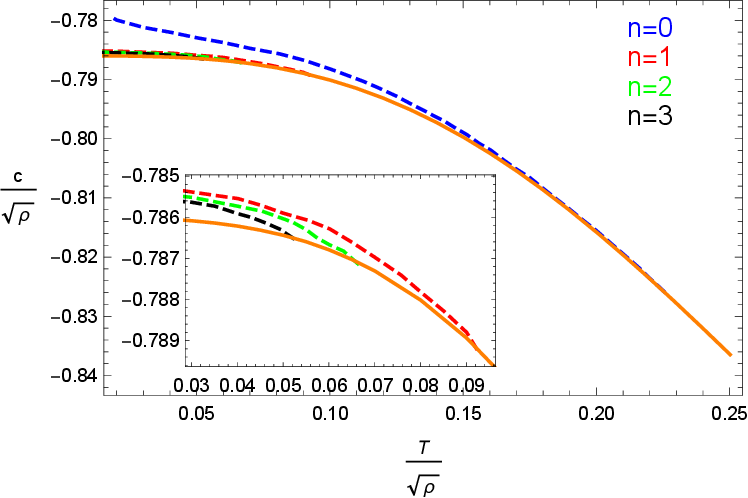}\hspace{0.4cm}%
\caption{ The HEE and HSC of scalar operators $\mathcal{O}_{+}$ (left) and $\mathcal{O}_{-}$ (right) with excited states versus temperature for the fixed width $l\sqrt{\rho}=1$, which shows that the higher excited state leads to a lager value of the HEE but a smaller value of the HSC for a given temperature in the superconducting phase. In each panel, the blue, red, green and black dashed lines denote the ground $(n=0)$, first $(n=1)$, second $(n=2)$ and third $(n=3)$ states of the superconducting phase, respectively. The orange solid line denotes the normal phase.}\label{4dHEE}
\end{figure}

In Fig. \ref{4dHEE}, we plot the HEE (top) and HSC (bottom) in terms of the temperature $T$ for operators $\mathcal{O}_{+}$ and $\mathcal{O}_{-}$ with excited states, respectively. In each panel, we can see that the HEE and HSC change discontinuously at the critical point where the curves of the normal phase intersect with those of the superconducting phase. It characterizes that, below the critical temperature, the system experiences a phase transition from a normal phase to a superconducting phase
with the decrease of the temperature. Moreover, both for the operator $\mathcal{O}_{+}$ and operator $\mathcal{O}_{-}$, the discontinuous points of the curves for the HEE and HSC corresponding to the critical temperatures from the ground state to the third excited state are given in Table \ref{4detemp}. It is obvious that the increase of the number of nodes $n$ makes the critical temperature $T_{c}$ of phase transitions decrease. On the other hand, there are some differences between the HEE and HSC. Firstly, the HEE increases as the temperature increases, and its value in the normal phase is larger than that in the superconducting phase. On the contrary, the HSC decreases with the increasing temperature and always has a smaller value in the normal phase than that in the superconducting phase. Secondly, it is interesting to find that, for a fixed temperature $T$, the higher excited state has a larger HEE but a smaller HSC in the superconducting phase.

\section{Entanglement entropy and complexity for excited states of holographic superconductors in 4-dimensional Einstein-Gauss-Bonnet gravity}

\subsection{Holographic model and condensates of the scalar field}

Now we study the fully backreacting holographic superconductor in the consistent $d\rightarrow 4$ Einstein-Gauss-Bonnet gravity \cite{Aoki}. In the Arnowitt-Deser-Misner (ADM) formalism, we adopt the metric ansatz
\begin{eqnarray}
ds^{2}=g_{\mu\nu}dx^{\mu}dx^{\nu}=-N^{2}dt^{2}+\gamma_{ij}(dx^{i}+N^{i}dt)(dx^{j}+N^{j}dt),
\end{eqnarray}
where $N$ is the lapse function, $\gamma_{ij}$ is the spatial metric and $N^{i}$ is the shift vector. We begin with a Maxwell field and a charged complex scalar field coupled via the action
\begin{eqnarray}\label{action4DEGB}
S=\int dtd^{3}xN\sqrt{\gamma}\bigg(\mathcal{L}_{EGB}^{4D}-\frac{1}{4}F_{\mu\nu}F^{\mu\nu}-|\nabla\psi-iqA\psi|^{2}-m^{2}|\psi|^{2}\bigg),
\end{eqnarray}
where the Lagrangian density reads
\begin{eqnarray}
\mathcal{L}_{EGB}^{4D}=\frac{1}{2\kappa^{2}}\bigg\{2R+\frac{6}{L^{2}}-\mathcal{M}+\frac{\alpha}{2}\bigg[8R^{2}-4R\mathcal{M}-\mathcal{M}^{2}-
\frac{8}{3}\bigg(8R_{ij}R^{ij}-4R_{ij}\mathcal{M}^{ij}-\mathcal{M}_{ij}\mathcal{M}^{ij}\bigg)\bigg]\bigg\},
\end{eqnarray}
with the Gauss-Bonnet coupling $\alpha$ and the Ricci tensor of the spatial metric $R_{ij}$. Here, we have
\begin{eqnarray}
\mathcal{M}_{ij}=R_{ij}+\mathcal{K}^{\kappa}_{\kappa}\mathcal{K}_{ij}-\mathcal{K}_{i\kappa}\mathcal{K}^{\kappa}_{j}, ~~~~~~\mathcal{M}\equiv\mathcal{M}^{i}_{i},
\end{eqnarray}
where $\mathcal{K}_{ij} \equiv  \left [\dot{\gamma}_{ij}-2D_{(i}N_{j)}-\gamma_{ij}D^2 \lambda_{\rm
GF} \right ]/(2N)$ with a dot denoting the derivative with respect to the time $t$, and $D_{i}$ being the covariant derivative compatible with the spatial metric.

We simply take the following ansatz of the metric
\begin{eqnarray}\label{4Dmetric}
N=\sqrt{f(r)}e^{-\chi(r)/2},~~~~~~N^{i}=0,~~~~~~\gamma_{ij}=diag\bigg(\frac{1}{f(r)},r^{2},r^{2}\bigg),
\end{eqnarray}
and consider the matter fields to be real functions of $r$, i.e., $\psi=|\psi(r)|$ and $A_{t}=\phi(r)$. So the equations of motion are
\begin{eqnarray}\label{4DGBchieq}
\chi^{\prime}+\frac{2\kappa^{2}r^{3}}{r^{2}-2\alpha f}\bigg(\psi'^2+\frac{q^{2}e^{\chi}\phi^{2}\psi^{2}}{f^{2}}\bigg)=0,
\end{eqnarray}
\begin{eqnarray}\label{4DGBfeq}
f^{\prime}-\frac{1}{r^{2}-2\alpha f}\bigg(\frac{3r^{3}}{L^{2}}-rf-\frac{\alpha f^{2}}{r}\bigg)+\frac{\kappa^{2}r^{3}}{r^{2}-2\alpha f}\bigg[m^{2}\psi^{2}+\frac{e^{\chi}\phi'^2}{2}+f\bigg(\psi'^2+\frac{q^{2}e^{\chi}\phi^{2}\psi^{2}}{f^{2}}\bigg)\bigg]=0,
\end{eqnarray}
\begin{eqnarray}\label{4Dphi}
\phi^{\prime\prime}+\bigg(\frac{2}{r}+\frac{\chi^{\prime}}{2}\bigg)\phi^{\prime}-\frac{2q^{2}\psi^{2}}{f}\phi=0,
\end{eqnarray}
\begin{eqnarray}\label{4Dpsi}
\psi^{\prime\prime}+\bigg(\frac{2}{r}+\frac{f^{\prime}}{f}-\frac{\chi^{\prime}}{2}\bigg)\psi^{\prime}+
\bigg(\frac{q^{2}e^{\chi}\phi^{2}}{f^{2}}-\frac{m^{2}}{f}\bigg)\psi=0,
\end{eqnarray}
where the prime denotes differentiation in $r$. If the Gauss-Bonnet term $\alpha\rightarrow 0$, Eqs. (\ref{4DGBchieq})-(\ref{4Dpsi}) will reduce to Eqs. (\ref{chieq})-(\ref{psi}) with $d=4$ for the backreacting holographic superconductors investigated in Ref. \cite{PanQY2012}. Here, the Hawking temperature has the same form as in (\ref{TH}), which is interpreted as the temperature of the dual field theory.

For the normal phase, we can get the analytic solution to the field equations (\ref{4DGBfeq}) and (\ref{4Dphi})
\begin{eqnarray}
f(r)&=&\frac{r^{2}}{2\alpha}\bigg[1-\sqrt{1-\frac{4\alpha}{L^{2}}\bigg(1-\frac{r_{+}^{3}}{r^{3}}\bigg)+\frac{2\alpha\kappa^{2}\rho^{2}}{r_{+}r^{3}}\bigg(1-\frac{r_{+}}{r}\bigg)}\bigg],\nonumber \\
\phi(r)&=&\mu-\frac{\rho}{r},
\end{eqnarray}
which reduces to the case of the $4$-dimensional AdS RN black hole in the limit $\alpha\rightarrow 0$.

For the superconducting phase, the boundary conditions at the horizon and asymptotic AdS boundary have to be imposed to solve Eqs. (\ref{4DGBchieq})-(\ref{4Dpsi}). At the horizon $r=r_{+}$, we still have the regularity conditions, just as in section II for the Einstein gravity. Near the asymptotic boundary $r\rightarrow \infty$, we find
\begin{eqnarray}
\chi\rightarrow0,~~f\sim\frac{r^{2}}{L_{eff}^{2}},~~\phi\sim\mu-\frac{\rho}{r},
~~\psi\sim\frac{\psi_{-}}{r^{\lambda_{-}}}+\frac{\psi_{+}}{r^{\lambda_{+}}},
\end{eqnarray}
where the effective AdS radius is defined as
\begin{eqnarray}
L_{eff}^{2}=\frac{2\alpha}{1-\sqrt{1-\frac{4\alpha}{L^{2}}}},
\end{eqnarray}
with the characteristic exponents $\lambda_{\pm}=(3\pm\sqrt{9+4m^{2}L_{eff}^{2}})/2$. Following Ref. \cite{QiaoXY}, we take $m^{2}L_{eff}^{2}=-2$, $\kappa=0.05$ and $\alpha\leq L^{2}/4$ (this is the so-called Chern-Simons limit) in our calculations. For simplicity, here we only consider the scalar operator $\mathcal{O}_{+}$ since the behaviors of the HEE and the HSC for both operators $\mathcal{O}_{+}$ and $\mathcal{O}_{-}$ are the same, just as shown in Fig. \ref{4dHEE} for the Einstein gravity.

\begin{figure}[ht]
\includegraphics[scale=0.41]{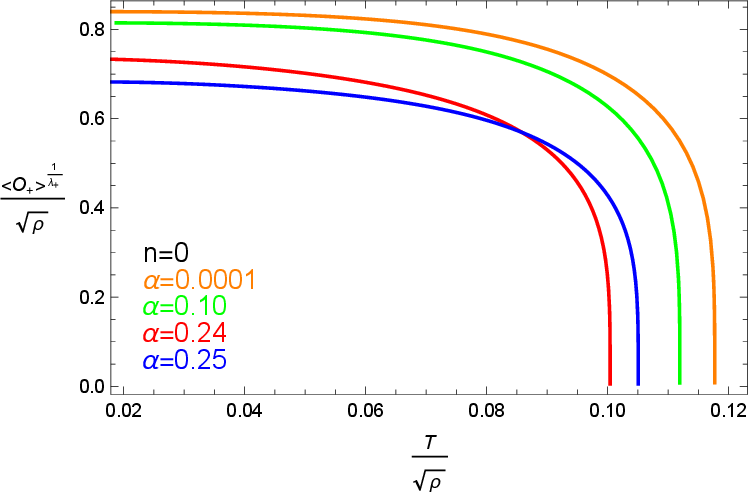}\hspace{0.4cm}%
\includegraphics[scale=0.41]{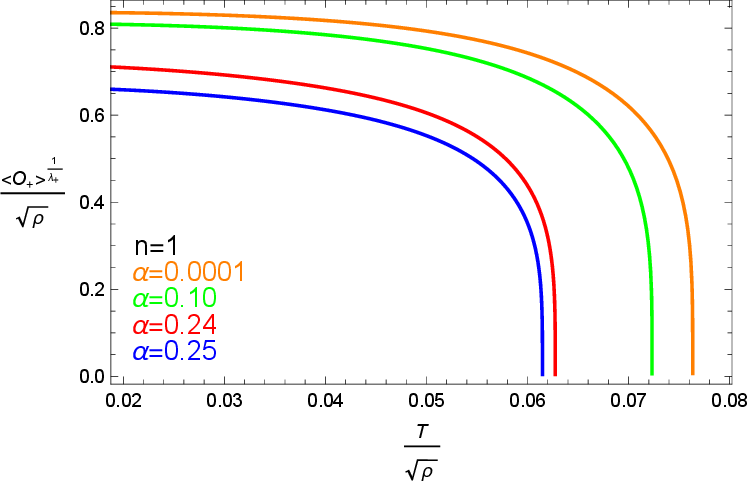}\hspace{0.4cm}%
\includegraphics[scale=0.41]{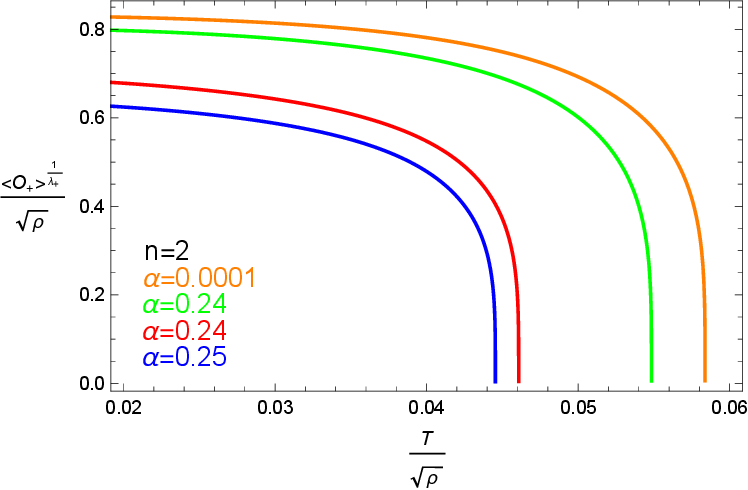}\hspace{0.4cm}%
\caption{The condensates of the scalar operator $\mathcal{O}_{+}$ versus temperature with the fixed mass $m^{2}L_{eff}^{2}=-2$ for different Gauss-Bonnet parameters $\alpha$, i.e., $\alpha=0.0001$ (orange), $0.10$ (green), $0.24$ (red) and $0.25$ (blue). The three panels from left to right represent the ground ($n=0$), first ($n=1$) and second ($n=2$) states, respectively. }\label{4Dcond}
\end{figure}

In Fig. \ref{4Dcond}, we present the scalar condensation $\langle O_{+}\rangle$ in terms of the temperature, by taking $\alpha=0.0001$, $0.10$, $0.24$ and $0.25$, for the ground $(n=0)$, first $(n=1)$ and second $(n=2)$ states. It is observed that the condensate occurs, for different values of $\alpha$ and $n$, if $T<T_{c}$. For small condensate, we see that there is a square root behavior $\langle O_{+}\rangle\sim(1-T/T_{c})^{1/2}$, which shows that the phase transition of these backreacting Gauss-Bonnet superconductors is typically second order one with the mean field critical exponent $1/2$ for all values of $\alpha$.

\begin{figure}[ht]
\includegraphics[scale=0.80]{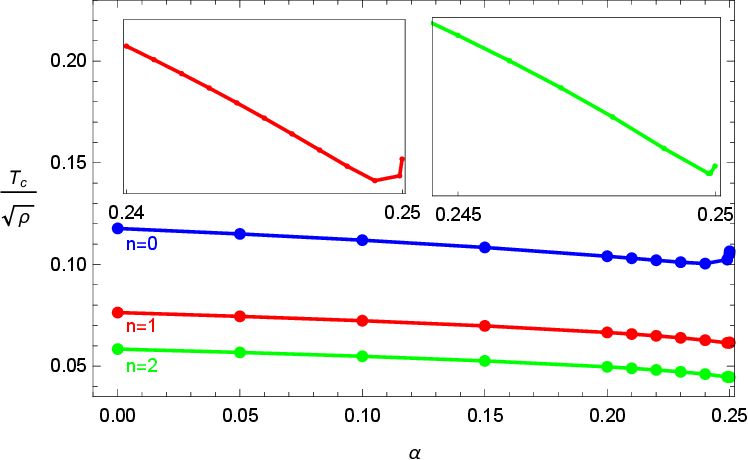}\hspace{0.4cm}%
\caption{The critical temperature $T_{c}$ of the scalar operator $\mathcal{O}_{+}$ versus Gauss-Bonnet parameter $\alpha$ with the fixed mass $m^{2}L_{eff}^{2}=-2$ for the ground $n=0$ (blue), first $n=1$ (red) and second $n=2$ (green) states, respectively.}\label{4Dtemp}
\end{figure}

In order to show the effect of the curvature correction on $T_{c}$, in Fig. \ref{4Dtemp} we present the critical temperature $T_{c}$ for the operator $\mathcal{O}_{+}$ in terms of the Gauss-Bonnet parameter $\alpha$, by taking $m^{2}L_{eff}^{2}=-2$, both for the ground and excited states. An interesting feature we observed is that the critical temperature $T_{c}$ decreases as $\alpha$ increases, but slightly increases near the Chern-Simons limit $\alpha=0.25$. Furthermore, this non-monotonic behavior of $T_{c}$ is more pronounced in the ground state than that in the excited state, which is in good agreement with the results obtained in Ref. \cite{PanJie}.

\subsection{HEE and HSC in the holographic model}

We are ready to analyze the properties of the HEE and HSC for the backreacting holographic superconductor in the $4$-dimensional Einstein-Gauss-Bonnet gravity. Since a Gauss-Bonnet term is present, we have to use a general formula to calculate the HEE \cite{Dong2014,Wald1993,Wald1995,Jacobson1994}
\begin{eqnarray}
\mathcal{S}=-2\pi\int d^{d}y\sqrt{-g}\bigg\{\frac{\partial \mathcal{L}}{\partial R_{\mu\rho\nu\sigma}}\varepsilon_{\mu\rho}\varepsilon_{\nu\sigma}-\sum_{\alpha}\bigg(\frac{\partial^{2}\mathcal{L}}{\partial R_{\mu_{1}\rho_{1}\nu_{1}\sigma_{1}}\partial R_{\mu_{2}\rho_{2}\nu_{2}\sigma_{2}}}\bigg)_{\alpha}\frac{2K_{\lambda_{1}\rho_{1}\sigma_{1}}K_{\lambda_{2}\rho_{2}\sigma_{2}}}{q_{\alpha}+1}\times \nonumber\\
\bigg[(n_{\mu_{1}\mu_{2}}n_{\nu_{1}\nu_{2}}-\varepsilon_{\mu_{1}\mu_{2}}\varepsilon_{\nu_{1}\nu_{2}})n^{\lambda_{1}\lambda_{2}}+(n_{\mu_{1}\mu_{2}}\varepsilon_{\nu_{1}\nu_{2}}+\varepsilon_{\mu_{1}\mu_{2}}n_{\nu_{1}\nu_{2}})\varepsilon^{\lambda_{1}\lambda_{2}}\bigg]\bigg\},
\end{eqnarray}
where $n_{\mu\nu}$ and $\varepsilon_{\mu\nu}$ reduce to the metric and Levi-Civita tensor in the two orthogonal directions when all other components vanish, and $q_{\alpha}$ is treated as ``anomaly coefficients". This will result in the corrections to expressions (\ref{RTformula}) and (\ref{HEE}). We again employ the shooting method in numerical calculations, and use $s$ and $c$ to denote the entanglement entropy and complexity of the universal term, respectively.

\begin{figure}[ht]
\includegraphics[scale=0.41]{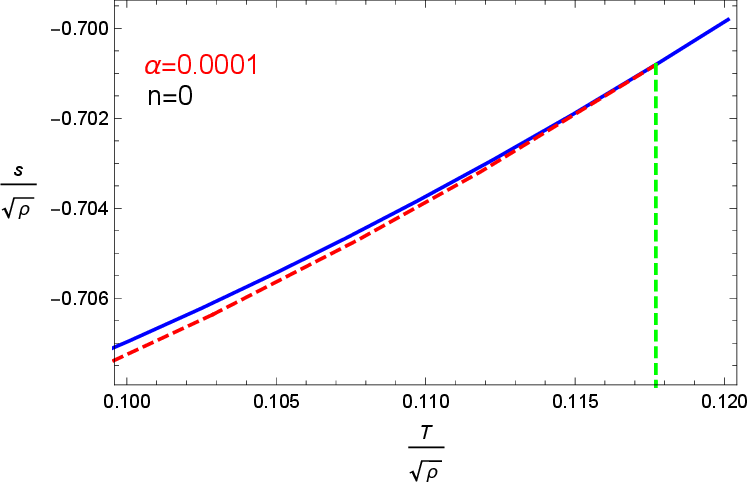}\hspace{0.4cm}%
\includegraphics[scale=0.41]{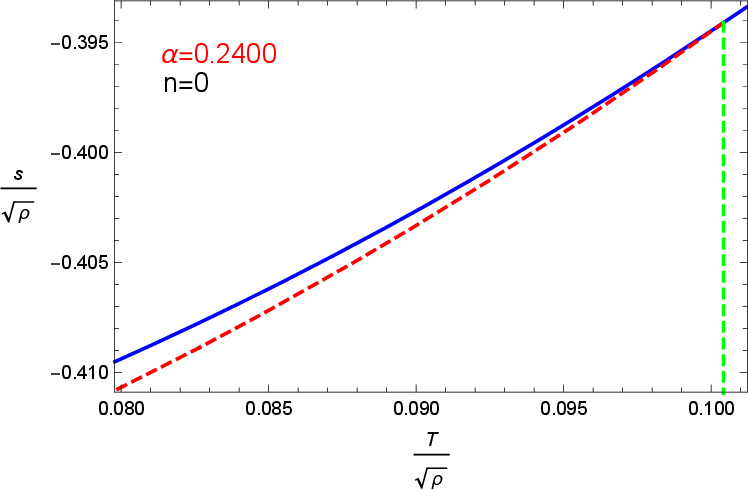}\hspace{0.4cm}%
\includegraphics[scale=0.41]{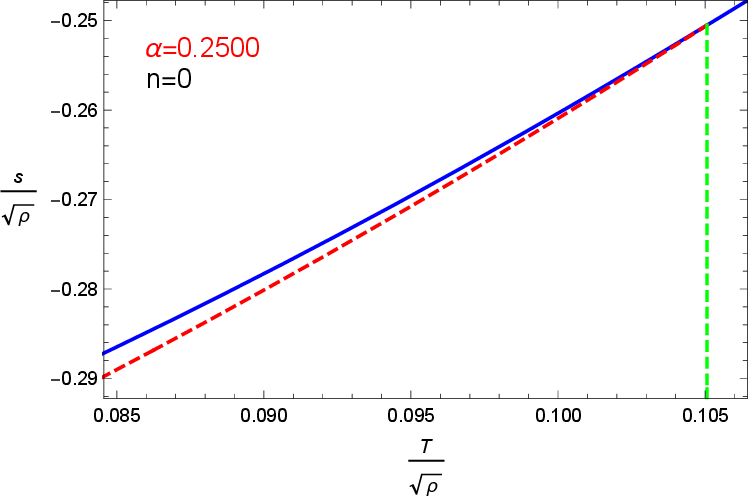}\hspace{0.4cm}%
\includegraphics[scale=0.41]{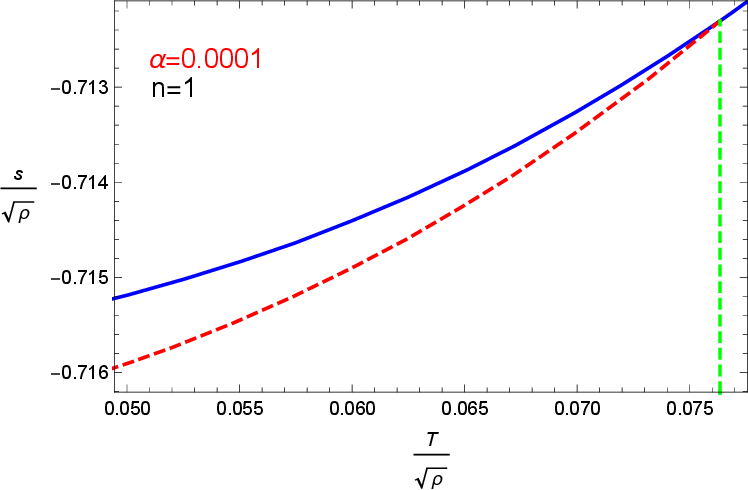}\hspace{0.4cm}%
\includegraphics[scale=0.41]{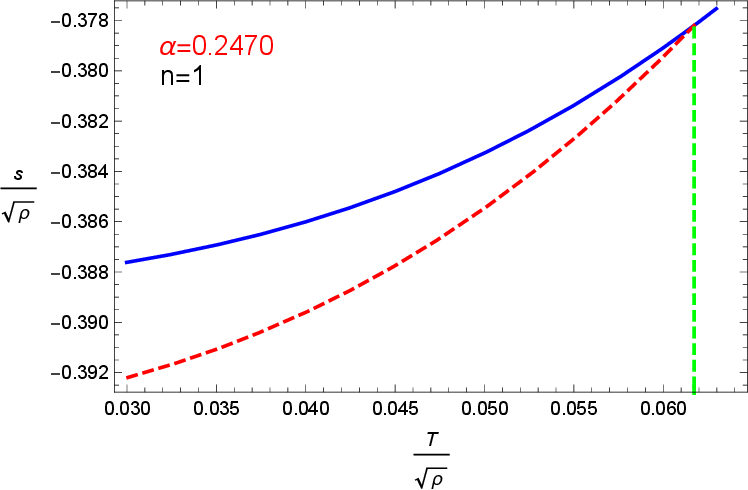}\hspace{0.4cm}%
\includegraphics[scale=0.41]{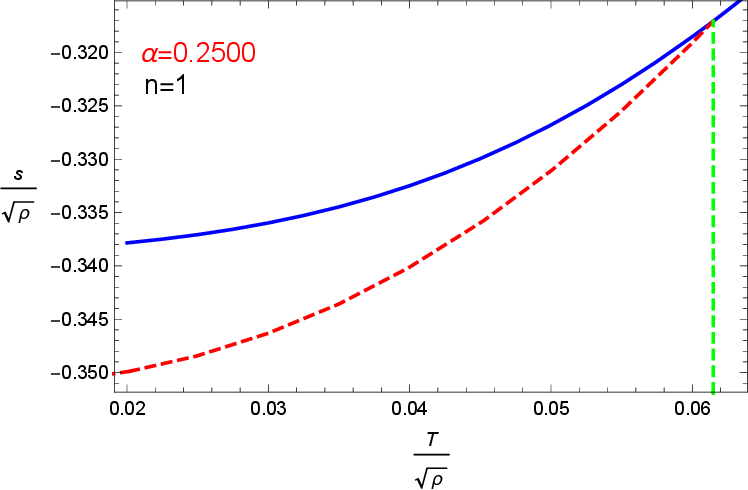}\hspace{0.4cm}%
\includegraphics[scale=0.41]{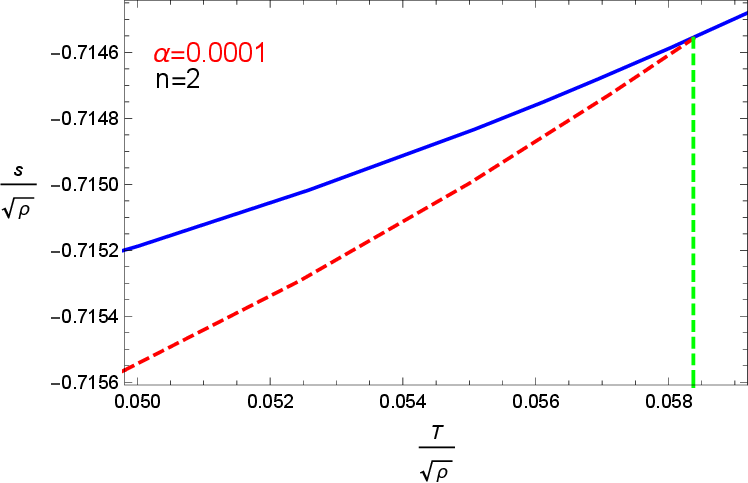}\hspace{0.4cm}%
\includegraphics[scale=0.41]{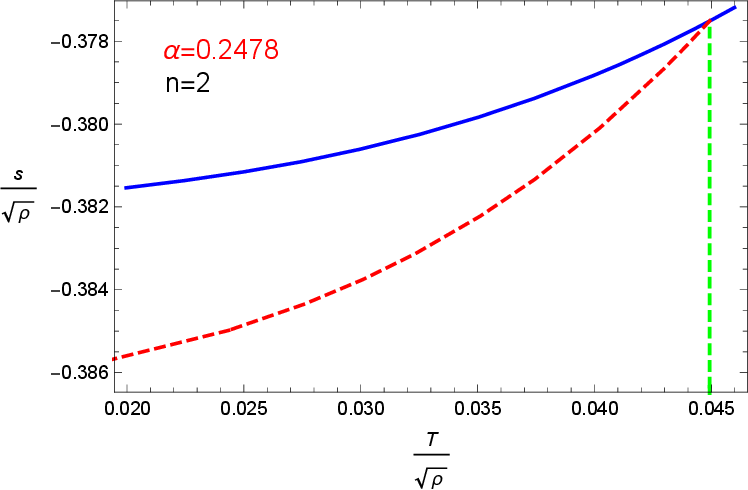}\hspace{0.4cm}%
\includegraphics[scale=0.41]{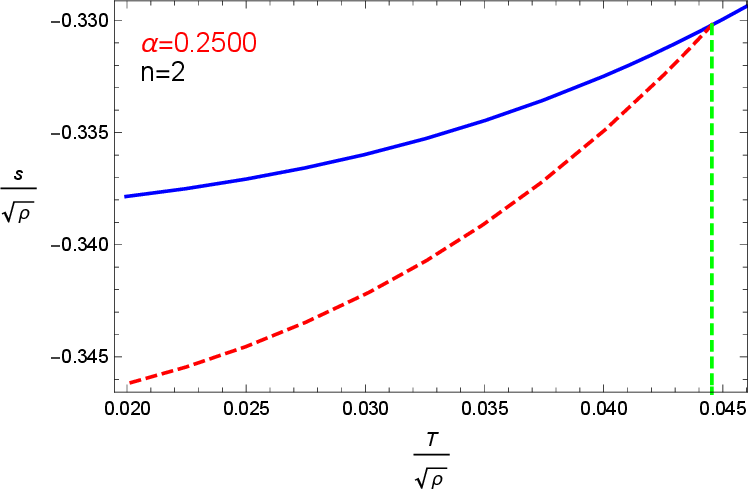}\hspace{0.4cm}%
\caption{ The HEE of the scalar operator $\mathcal{O}_{+}$ versus temperature from the ground state ($n=0$) to the second excited state ($n=2$) with a fixed width $l\sqrt{\rho}=1$ for different Gauss-Bonnet parameters $\alpha$, which shows that the HEE always increases monotonously with the increase of the temperature and its value in the normal phase always larger than that in the superconducting phase. The solid (blue) lines denote the normal phase and the dashed (red) lines are for the superconducting phase. }\label{4DZHEE}
\end{figure}

In Fig. \ref{4DZHEE}, we present the HEE of the scalar operator $\mathcal{O}_{+}$ in terms of the temperature $T$, for both the ground state $(n=0)$ and up to the second excited state $(n=2)$. In each panel, the critical temperature $T_{c}$ for the system can be determined by the joint point of the solid line for the normal phase and the dashed line for the superconducting phase. For example, in the ground state $(n=0)$, we have $T_{c}/\sqrt{\rho}=0.117705$ for $\alpha=0.0001$ (top-left panel), $T_{c}/\sqrt{\rho}=0.100418$ for $\alpha=0.24$ (top-middle panel) and $T_{c}/\sqrt{\rho}=0.105064$ for $\alpha=0.25$ (top-right panel), which are consistent with those in Fig. \ref{4Dtemp}. It is shown obviously that, for the ground state and as $\alpha$ approaches the Chern-Simons limit, the critical temperature first decreases and then increases. In addition, for the ground state and excited states, we can always find that the value of the HEE in the superconducting phase is less than that in the normal phase when $T<T_{c}$, which does not depend on the Gauss-Bonnet parameter $\alpha$. This behavior of the HEE is due to the fact that the condensate turns on at the critical point $T_{c}$ and the formation of Cooper pairs makes the degrees of freedom decrease in the superconducting phase. While we fix the Gauss-Bonnet parameter $\alpha$, we see that the value of the HEE becomes larger as the number of nodes $n$ increases.

\begin{figure}[ht]
\includegraphics[scale=0.41]{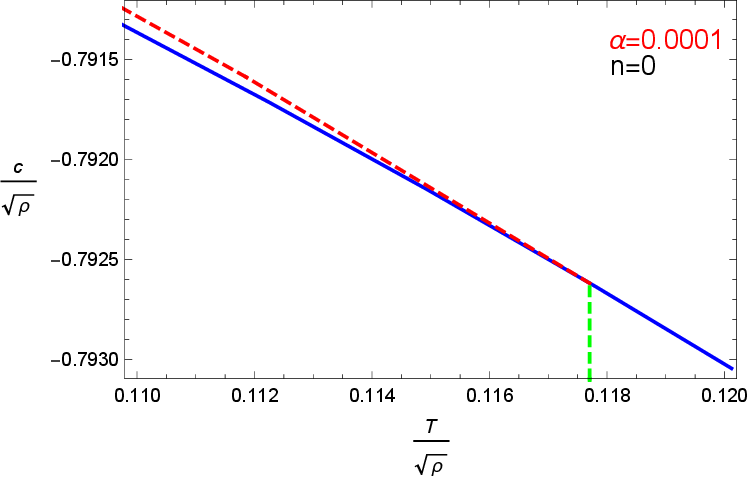}\hspace{0.4cm}%
\includegraphics[scale=0.41]{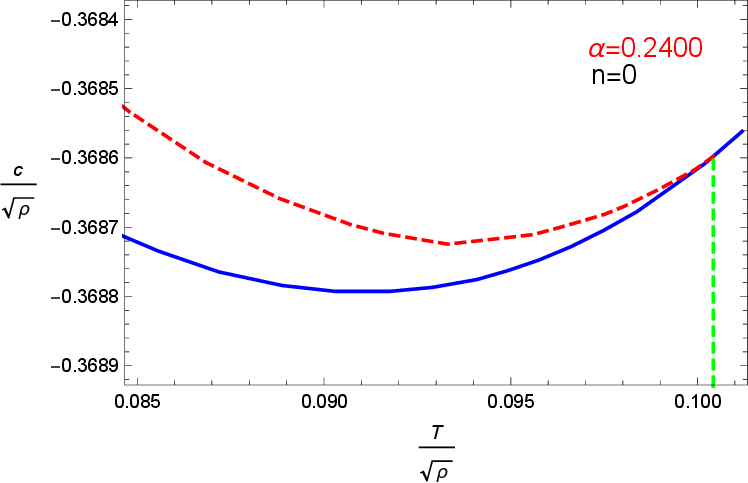}\hspace{0.4cm}%
\includegraphics[scale=0.41]{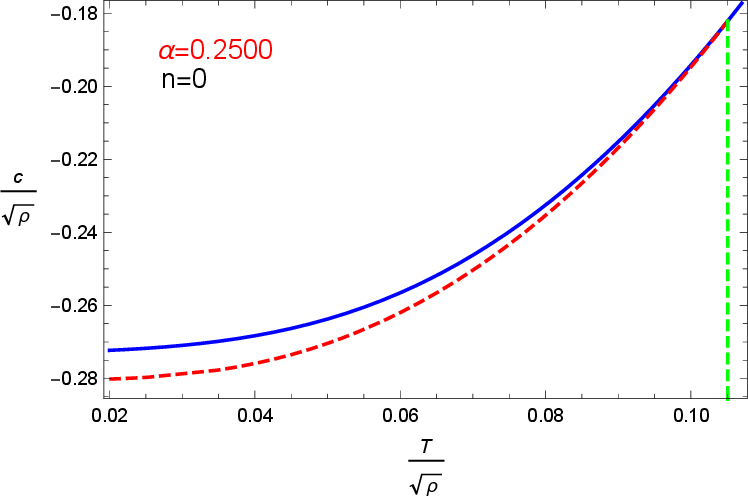}\hspace{0.4cm}%
\includegraphics[scale=0.41]{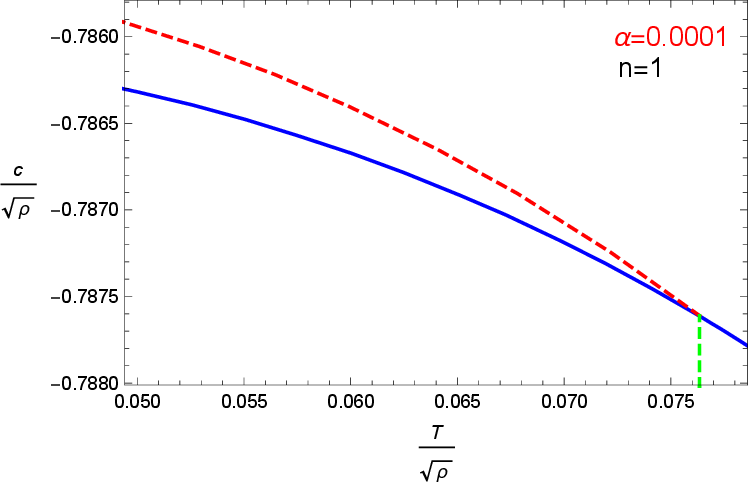}\hspace{0.4cm}%
\includegraphics[scale=0.41]{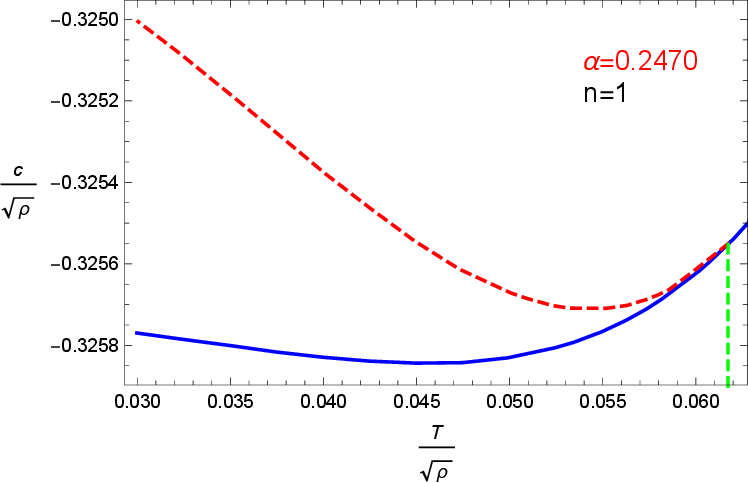}\hspace{0.4cm}%
\includegraphics[scale=0.41]{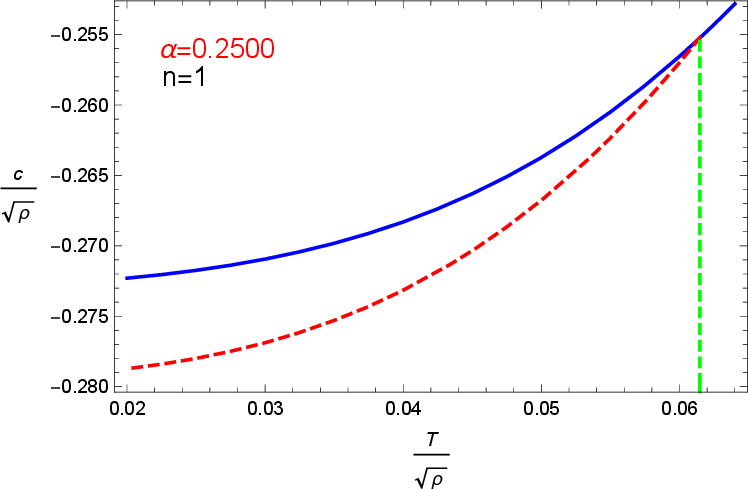}\hspace{0.4cm}%
\includegraphics[scale=0.41]{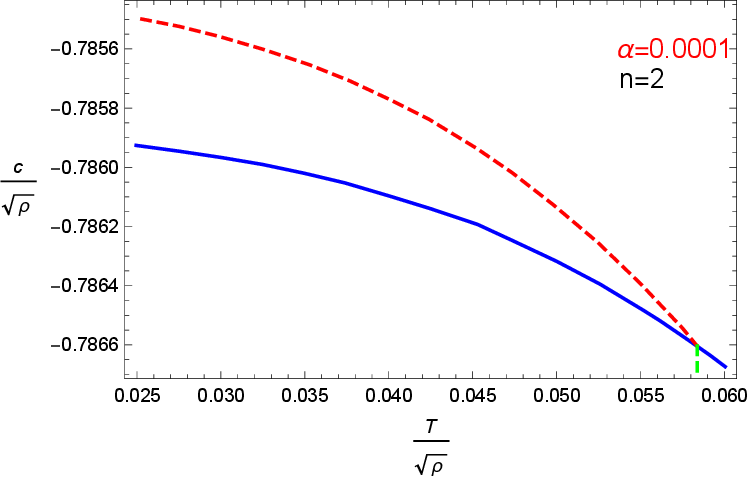}\hspace{0.4cm}%
\includegraphics[scale=0.41]{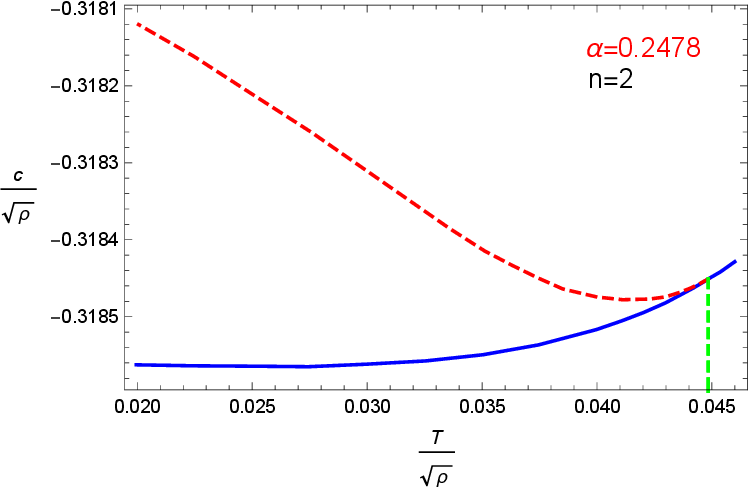}\hspace{0.4cm}%
\includegraphics[scale=0.41]{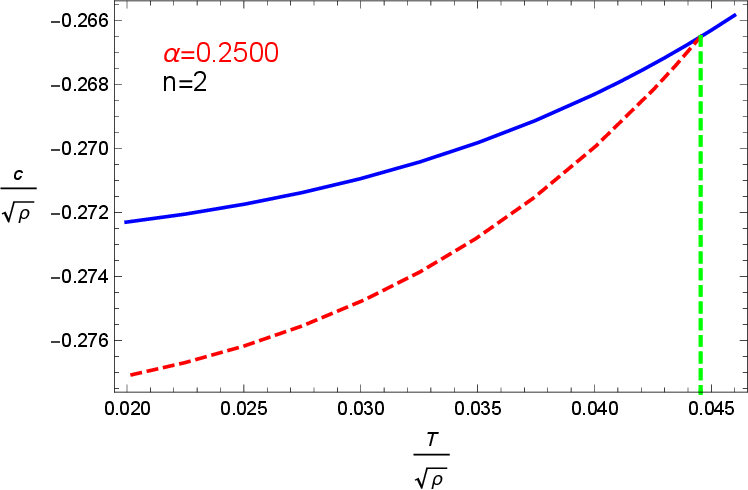}\hspace{0.4cm}%
\caption{ The HSC of the scalar operator $\mathcal{O}_{+}$ versus temperature from the ground state ($n=0$) to the second excited state ($n=2$) with a fixed width $l\sqrt{\rho}=1$ for different Gauss-Bonnet parameters $\alpha$, which shows that
the Gauss-Bonnet parameter has a more subtle effect on the HSC when compared to the HEE. The solid (blue) lines denote the normal phase and the dashed (red) lines are for the superconducting phase. }\label{4DZHSC}
\end{figure}

In Fig. \ref{4DZHSC}, we plot the HSC of the scalar operator $\mathcal{O}_{+}$ in terms of the temperature $T$, and find that the curve of the HSC in the normal phase and the one in the superconducting phase intersect at the same critical temperature as that reflected by the HEE in Fig. \ref{4DZHEE}. It means that the HSC is able to capture the emergence of the phase transitions in the ground state and excited states. What is noteworthy is that the Gauss-Bonnet term has an interesting effect on the relation between the HSC and the temperature, which can be seen in the ground state and excited states. Obviously, there is a threshold $\alpha_t$ of the Gauss-Bonnet parameter. When $\alpha<\alpha_t$, the value of the HSC decreases as the temperature increases. And the value of the HSC decreases as $n$ increases for the fixed $\alpha$, which is consistent with the finding obtained from the bottom-left panel of Fig. \ref{4dHEE}. At the threshold $\alpha=\alpha_t$, with the increasing $T/\sqrt{\rho}$, the value of the HSC first decreases and then increases. It should be noted that the threshold becomes larger with the higher excited state, i.e., $\alpha_t=0.2400$ for the ground state $n=0$, $\alpha_t=0.2470$ for the first state $n=1$ and $\alpha_t=0.2478$ for the second state $n=2$. Whereas as $\alpha$ goes up to the Chern-Simons limit $\alpha=0.25$, this non-monotonic behavior of the HSC will convert to a monotonic increasing function of $T/\sqrt{\rho}$, which is contrary to the case of $\alpha<\alpha_t$. Besides, we find that the normal phase always has a smaller HSC than the superconducting phase except for the situation of the Chern-Simons limit, namely, the value of the HSC in the normal phase is larger than that in the superconducting phase for $\alpha=0.25$. Under the influence of the Gauss-Bonnet parameter, these special features of the HSC with respect to the temperature imply that the curvature correction makes a difference to the properties of the spacetime and changes the geometric structure.

\begin{figure}[ht]
\includegraphics[scale=0.41]{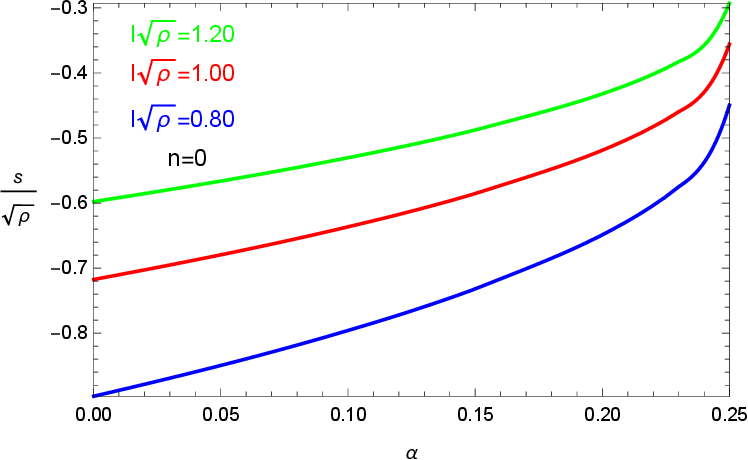}\hspace{0.4cm}%
\includegraphics[scale=0.41]{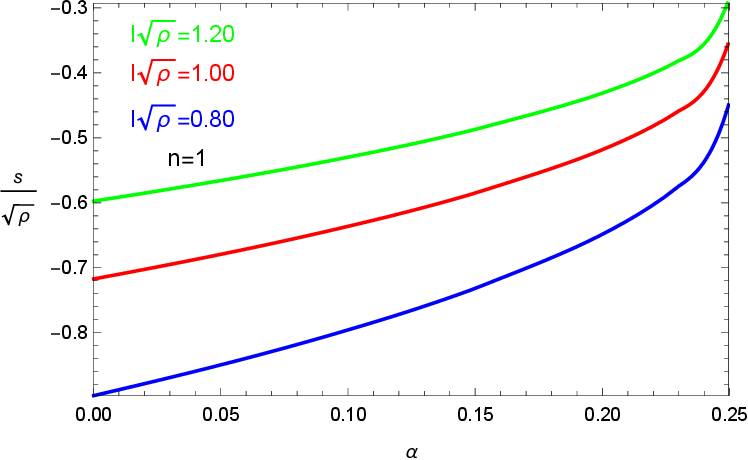}\hspace{0.4cm}%
\includegraphics[scale=0.41]{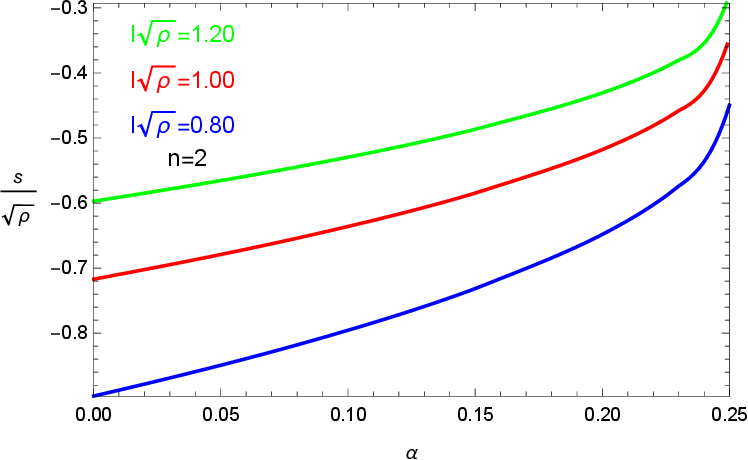}\hspace{0.4cm}%
\includegraphics[scale=0.41]{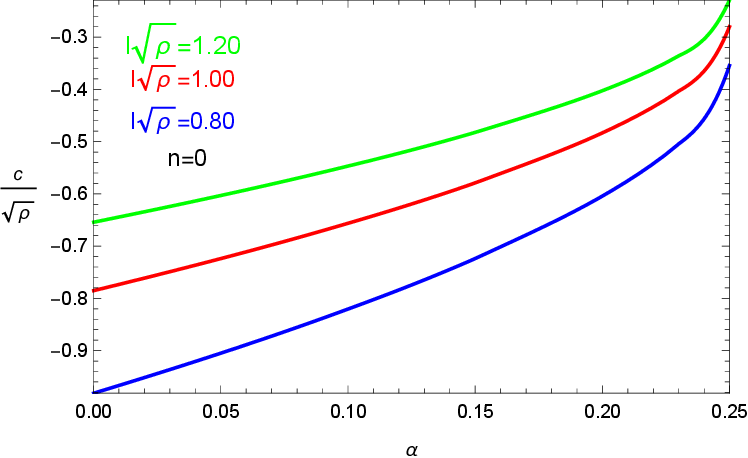}\hspace{0.4cm}%
\includegraphics[scale=0.41]{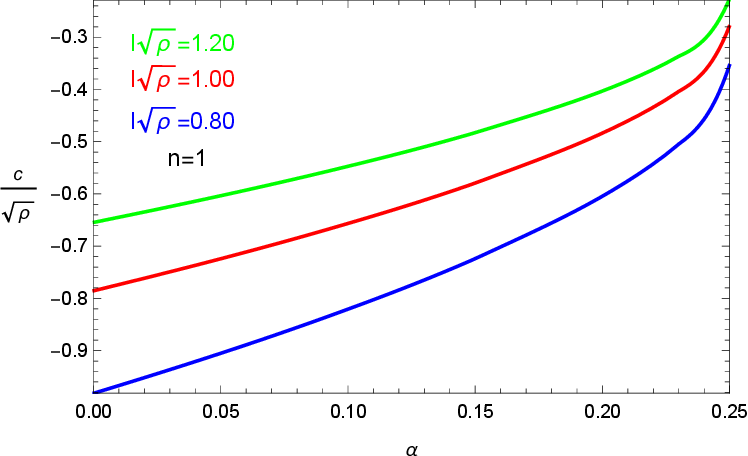}\hspace{0.4cm}%
\includegraphics[scale=0.41]{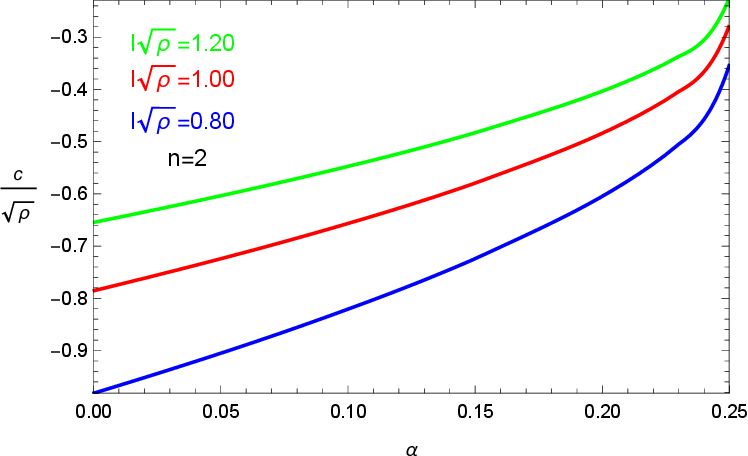}\hspace{0.4cm}%
\caption{ The HEE and HSC of the scalar operator $\mathcal{O}_{+}$ versus Gauss-Bonnet parameter $\alpha$ at the temperature $T/\sqrt{\rho}=0.02$ with some chosen values of the widths, i.e., $l\sqrt{\rho}=0.80$ (blue), $l\sqrt{\rho}=1.00$ (red) and $l\sqrt{\rho}=1.20$ (green). The panels from left to right represent the ground ($n=0$), first ($n=1$) and second ($n=2$) states, respectively.}\label{4DZHEEHSCalpha}
\end{figure}

On the other hand, for the fixed temperature, from Figs. \ref{4DZHEE} and \ref{4DZHSC} we note that the larger Gauss-Bonnet parameter $\alpha$ leads to the larger values of the HEE and HSC. To further illustrate this, in Fig. \ref{4DZHEEHSCalpha}, we plot the HEE and HSC versus the Gauss-Bonnet parameter $\alpha$, respectively, from the ground state to the second excited state at a fixed temperature $T/\sqrt{\rho}=0.02$ with different widths, i.e., $l\sqrt{\rho}=0.80$, $l\sqrt{\rho}=1.00$ and $l\sqrt{\rho}=1.20$. We observe clearly that, regardless of the width and the number of nodes, the HEE and HSC increase with the increase of the Gauss-Bonnet correction. Moreover, in each panel, the HEE and HSC become larger as the width increases.

\section{Conclusion}

In this work, we first study the HEE and HSC for the excited states of holographic superconductors with full backreaction in the $4$-dimensional Einstein gravity. We note that the changes of the HEE and HSC both for the scalar operators $\mathcal{O}_{-}$ and $\mathcal{O}_{+}$ are discontinuous at the critical temperature $T_{c}$, and $T_{c}$ in the excited states is lower than that in the ground state, which indicates that the higher excited state makes the scalar condensate harder to form. The values of $T_{c}$ reflected by the HEE and HSC are consistent with the results obtained from the condensate behavior, which means that both the HEE and HSC can be utilized as good probes to the superconducting phase transition in the excited state. However, there are some differences between the HEE and HSC. We observe that, for the ground state and excited states, the value of the HEE in the normal phase is larger than that in the superconducting phase and increases as the temperature increases, which is the opposite to the behavior of the HSC, namely, the normal phase always has a smaller HSC than the superconducting phase and the HSC decreases as the temperature increases. Meanwhile, with a fixed temperature $T$ in the superconducting phase, the higher excited state leads to a lager value of the HEE but a smaller value of the HSC.

Next, we extend the investigation on the HEE and HSC for the excited states of backreacting superconductors to the $4$-dimensional Einstein-Gauss-Bonnet gravity. One remarkable feature is that, for the scalar operator $\mathcal{O}_{+}$, the critical temperature $T_{c}$ decreases as the higher curvature correction $\alpha$ increases, but slightly increases as $\alpha$ grows to the Chern-Simons limit, and this non-monotonic bahavior of $T_{c}$ is more pronounced in the ground state than in the excited state, which can be supported by the findings obtained from both the HEE and HSC. On the other hand, regardless of $\alpha$, we find that the HEE always increases monotonously with the increase of the temperature and the superconducting phase always has a smaller HEE than the normal phase below $T_{c}$ both for the ground state and excited states. Furthermore, the higher excited state leads to a lager value of the HEE for a given temperature $T$ in the superconducting phase, regardless of the Einstein gravity or the Einstein-Gauss-Bonnet gravity. Obviously, these may be the general features for the HEE in the holographic superconductors. However, the story is completely different if we study the HSC for the excited states of holographic superconductors in the Einstein-Gauss-Bonnet gravity. The noteworthy feature is that the effect of $\alpha$ on the relation between the HSC and the temperature is nontrivial in the ground state and excited states, which is a distinguishing property of the HSC and has not been found in the HEE. Specifically, the HSC always behaves as a monotonic decreasing function of the temperature till $\alpha$ reaches some threshold $\alpha_t$. At this critical point $\alpha_t$, the HSC changes non-monotonously with the temperature, i.e., it first decreases and then increases with the increasing temperature. It is shown that the value of $\alpha_t$ will be closer to the Chern-Simons limit in the higher excited state. More interestingly, when $\alpha$ approaches to the Chern-Simons limit, the HSC will convert to a monotonic increasing function of the temperature. Besides, the value of the HSC in the normal phase is less than that in the superconducting phase for $\alpha\leq\alpha_t$, but it is just on the contrary for the Chern-Simons limit. Lastly, we find that, for the ground state and excited states, the increase of $\alpha$ makes both the HEE and HSC increase, which is independent of the strip width. Thus, we conclude that the HEE and HSC provide richer physics in the phase transition and scalar condensate for holographic superconductors with excited states.

\begin{acknowledgments}

We would like to acknowledge helpful discussions with Jian-Pin Wu and Guoyang Fu. This work was supported by the National Key Research and Development Program of China (Grant No. 2020YFC2201400), National Natural Science Foundation of China (Grant Nos. 12275079, 12035005 and 11690034) and Postgraduate Scientific Research Innovation Project of Hunan Province (Grant No. CX20210472).

\end{acknowledgments}


\begin{thebibliography}{99}

\bibitem{Maldacena}
J. Maldacena,
Adv. Theor. Math. Phys. {\bf 2}, 231 (1998) [Int. J. Theor. Phys. {\bf 38}, 1113 (1999)].

\bibitem{Witten}
E. Witten,
Adv. Theor. Math. Phys. {\bf 2}, 253 (1998).

\bibitem{Gubser}
S.S. Gubser, I.R. Klebanov, and A.M. Polyakov,
Phys. Lett. B {\bf 428}, 105 (1998).

\bibitem{HartnollPRL}
S.A. Hartnoll, C.P. Herzog, and G.T. Horowitz, 
Phys. Rev. Lett. {\bf 101}, 031601 (2008).

\bibitem{HartnollJHEP}
S.A. Hartnoll, C.P. Herzog, and G.T. Horowitz, 
J. High Energy Phys. {\bf 12}, 015 (2008).

\bibitem{Gubser-Pufu}
S.S. Gubser and S.S. Pufu, 
J. High Energy Phys. {\bf 11}, 033 (2008).

\bibitem{CaiPWave-1}
R.G. Cai, S. He, L. Li, and L.F. Li, 
J. High Energy Phys. {\bf 12}, 036 (2013).

\bibitem{DWaveChen}
J.W. Chen, Y.J. Kao, D. Maity, W.Y. Wen, and C.P. Yeh, 
Phys. Rev. D {\bf 81}, 106008 (2010).

\bibitem{DWaveBenini}
F. Benini, C.P. Herzog, R. Rahman, and A. Yarom, 
J. High Energy Phys. {\bf 11}, 137 (2010).

\bibitem{CaiQongGen2015}
R.G. Cai, L. Li, L.F. Li, and R.Q. Yang, 
Sci. China-Phys. Mech. Astron. {\bf 58}, 060401 (2015); arXiv:1502.00437 [hep-th].

\bibitem{Hartnoll2009}
S.A. Hartnoll,
Class. Quant. Grav. {\bf 26}, 224002 (2009).

\bibitem{Herzog2009}
C.P. Herzog, 
J. Phys. A {\bf 42}, 343001 (2009).

\bibitem{Horowitz2011}
G.T. Horowitz, 
Lect. Notes Phys. {\bf 828}, 313 (2011); arXiv:1002.1722 [hep-th].

\bibitem{Zharkov}
G.F. Zharkov, 
Phys. Rev. B {\bf 63}, 224513 (2001).

\bibitem{Peeters}
F. Peeters, V. Schweigert, B. Baelus, and P. Deo, 
Physica C {\bf 332}, 255 (2000).

\bibitem{Mooij}
J.E. Mooij and C.J.P. Harmans, 
N. J. Phys. {\bf 7}, 219 (2005).

\bibitem{Lyatti}
M. Lyatti, M.A. Wolff, I. Gundareva, M. Kruth, S. Ferrari, R.E. Dunin-Borkowski, and C. Schuck,
Nat. Commun. 11, 763 (2020)


\bibitem{Coffey}
D. Coffey, L.J. Sham, and Y.R. Lin-Liu, 
Phys. Rev. B {\bf 38}, 5084(R) (1988).

\bibitem{Sahoo}
S. Sahoo, 
Phys. Rev. B {\bf 60}, 10803 (1999).

\bibitem{Demler}
E. Demler, W. Hanke, and S.C. Zhang, 
Rev. Mod. Phys. {\bf 76}, 909 (2004).

\bibitem{Semenov}
A.V. Semenov, I.A. Devyatov, P.J. de Visser, and T.M. Klapwijk, 
Phys. Rev. Lett. {\bf 117}, 047002 (2016).

\bibitem{WangYQ}
Y.Q. Wang, T.T. Hu, Y.X. Liu, J. Yang, and L. Zhao, 
J. High Energy Phys. {\bf 06}, 013 (2020); arXiv:1910.07734 [hep-th].

\bibitem{Vodolazov}
D.Y. Vodolazov and F.M. Peeters, 
Phys. Rev. B {\bf 66}, 054537 (2002).

\bibitem{WangYQBackreaction}
Y.Q. Wang, H.B. Li, Y.X. Liu, and Y. Zhong, 
Eur. Phys. J. C  {\bf 81}, 628 (2021); arXiv:1911.04475 [hep-th].

\bibitem{QiaoXY2020}
X.Y. Qiao, D. Wang, L. OuYang, M.J. Wang, Q.Y. Pan, and J.L. Jing, 
Phys. Lett. B {\bf 811}, 135864 (2020); arXiv:2007.08857 [hep-th].

\bibitem{OuYangLiang2021}
L. Ouyang, D. Wang, X.Y. Qiao, M.J. Wang, Q.Y. Pan, and J.L. Jing,
Sci. China-Phys. Mech. Astron. {\bf 64}, 240411 (2021); arXiv:2010.10715 [hep-th].

\bibitem{Liran2020}
R. Li, J. Wang, Y.Q. Wang, and H.B. Zhang, 
J. High Energy Phys. {\bf 11}, 059 (2020); arXiv:2008.07311 [hep-th].

\bibitem{XiangZW}
Q. Xiang, L. Zhao, and Y.Q. Wang, 
arXiv:2010.03443 [hep-th].

\bibitem{PanJie}
J. Pan, X.Y. Qiao, D. Wang, Q.Y. Pan, Z.Y. Nie, and J.L. Jing,
Phys. Lett. B   {\bf 823}, 136755 (2021); arXiv:2109.02207 [hep-th].

\bibitem{YuBao}
Y. Bao, H. Guo, and X.M. Kuang,
Phys. Lett. B {\bf 822}, 136646 (2021).

\bibitem{ZhangZPJNPB}
S.H. Zhang, Z.X. Zhao, and Q.Y. Pan, and J.L. Jing,
Nucl. Phys. B {\bf 976}, 115701 (2022); arXiv:2107.09486 [hep-th].

\bibitem{Xiang2022}
Q. Xiang, L. Zhao, and Y.Q. Wang,
arXiv:2207.10593 [hep-th].


\bibitem{Ryu}
S. Ryu and T. Takayanagi, Phys. Rev. Lett.  {\bf 96}, 181602 (2006); arXiv:hep-th/0603001.

\bibitem{Ryu2006}
S. Ryu and T. Takayanagi, J. High Energy Phys.  {\bf 08}, 045 (2006); arXiv:hep-th/0605073.

\bibitem{Albash}
T. Albash and C.V. Johnson, J. High Energy Phys.  {\bf 05}, 079 (2012); arXiv:1202.2605 [hep-th].

\bibitem{Ogawa}
N. Ogawa and T. Takayanagi, J. High Energy Phys.  {\bf 10}, 147 (2011).

\bibitem{XiDong}
X. Dong, J. High Energy Phys.  {\bf 01}, 044 (2014); arXiv:1310.5713 [hep-th].

\bibitem{CaiRongGen027}
R.G. Cai, S. He, L. Li, and Y.L. Zhang, J. High Energy Phys.  {\bf 07}, 027 (2012).

\bibitem{LiLiFang}
L.F. Li, R.G. Cai, L. Li, and C. Shen,
Nucl. Phys. B  {\bf 894}, 15 (2015); arXiv:1310.6239 [hep-th].

\bibitem{CaiRongGen088}
R.G. Cai, S. He, L. Li and Y.L. Zhang, J. High Energy Phys. {\bf 07}, 088 (2012).

\bibitem{CaiRongGen107}
R.G. Cai, S. He, L. Li, and L.F. Li, J. High Energy Phys.  {\bf 10}, 107 (2012).

\bibitem{Kuang}
X.M. Kuang, E. Papantonopoulos, and B. Wang, J. High Energy Phys. {\bf 05}, 130 (2014).

\bibitem{Peng}
Y. Peng and Q.Y. Pan, J. High Energy Phys. {\bf 06}, 011 (2014).

\bibitem{YaoWeiping}
W.P Yao and J.L Jing, J. High Energy Phys. {\bf 05}, 058 (2014).

\bibitem{Jeong2022}
H.S. Jeong, K.Y. Kim, and Y.W. Sun,
J. High Energy Phys. {\bf 06}, 078 (2022); arXiv:2203.07612 [hep-th].

\bibitem{Susskind2014}
L. Susskind,
Fortsch. Phys. {\bf 64}, 49 (2016); arXiv:1411.0690 [hep-th].

\bibitem{SusskindCV2014}
D. Stanford and L. Susskind, Phys. Rev. D {\bf 90}, 126007 (2014); arXiv:1406.2678[hep-th].

\bibitem{SusskindCV2016}
L. Susskind, Fortschr. Phys. {\bf 64}, 24 (2016); arXiv:1402.5674[hep-th].

\bibitem{BrownCAprl}
A.R. Brown, D.A. Roberts, L. Susskind, B. Swingle, and Y. Zhao, Phys. Rev. Lett. {\bf 116}, 191301 (2016).

\bibitem{BrownCAprd}
A.R. Brown, D.A. Roberts, L. Susskind, B. Swingle, and Y. Zhao, Phys. Rev. D {\bf 93}, 086006 (2016).

\bibitem{Alishahiha}
M. Alishahiha, Phys. Rev. D {\bf 92}, 126009 (2015).

\bibitem{Momeni}
D. Momeni, S.A.H. Mansoori, and R. Myrzakulov, Phys. Lett. B  {\bf 756}, 354 (2016).

\bibitem{Zangeneh}
M.K. Zangeneh, Y.C. Ong, and B. Wang, Phys. Lett. B {\bf 771}, 235 (2017).

\bibitem{YangJNK}
R.Q. Yang, H.S. Jeong, C. Niu, and K.Y. Kim,
J. High Energy Phys. {\bf 04}, 146 (2019).

\bibitem{ChakrabortyCQG}
A. Chakraborty,
Class. Quant. Grav. {\bf 37}, 065021 (2020).



\bibitem{Fujita}
M. Fujita, Prog. Theor. Exp. Phys.  {\bf 063}, B04 (2019); arXiv:1810.09659 [hep-th].

\bibitem{Stuckelbergsuperconductor}
H. Guo, X.M. Kuang, and B. Wang, Phys. Lett. B {\bf 797}, 134879 (2019); arXiv:1902.07945[hep-th].

\bibitem{Shiyu2020}
Y. Shi, Q.Y. Pan, and J.L. Jing, Eur. Phys. J. C  {\bf 80}, 1100 (2020).

\bibitem{Lainonlinear2022}
C.Y. Lai and Q.Y. Pan, Nucl. Phys. B {\bf 974}, 115615 (2022).

\bibitem{YuSenAn}
Y.S. An, L. Li, F.G. Yang, and R.Q. Yang,
J. High Energy Phys. {\bf 08}, 133 (2022).


\bibitem{GB1}
B. Zwiebach, Phys. Lett. B {\bf 156}, 315 (1985).

\bibitem{GB2}
D.J. Gross and E. Witten, Nucl. Phys. B {\bf 277}, 1 (1986).

\bibitem{GB3}
D.J. Gross and J.H. Sloan, Nucl. Phys. B {\bf 291}, 41 (1987).

\bibitem{CaiRongGen2002}
R.G. Cai, Phys. Rev. D {\bf 65}, 084014 (2002); arXiv:hep-th/0109133.

\bibitem{Gregory}
R. Gregory, S. Kanno, and J. Soda, J. High Energy Phys. {\bf 10}, 010 (2009).

\bibitem{PanQiYuan2010}
Q.Y. Pan, B. Wang, E. Papantonopoulos, J. Oliveria, and A.B. Pavan, Phys. Rev. D {\bf 81}, 106007 (2010).

\bibitem{Gregory2011}
R. Gregory, J. Phys. Conf. Ser. {\bf 283}, 012016 (2011); arXiv:1012.1558 [hep-th].

\bibitem{Kanno}
S. Kanno, Class. Quant. Grav. {\bf 28}, 127001 (2011).

\bibitem{Gangopadhyay}
S. Gangopadhyay and D. Roychowdhury, J. High Energy Phys. {\bf 05}, 156 (2012).

\bibitem{Ghorai}
D. Ghorai and S. Gangopadhyay, Eur. Phys. J. C {\bf 76}, 146 (2016).

\bibitem{Sheykhi}
A. Sheykhi, H.R. Salahi, and A. Montakhab, J. High Energy Phys. {\bf 04}, 058 (2016).

\bibitem{Salahi}
H.R. Salahi, A. Sheykhi, and A. Montakhab, Eur. Phys. J. C {\bf 76}, 575 (2016); arXiv:1608.05025 [gr-qc].

\bibitem{LiZH}
Z.H. Li, Y.C. Fu, and Z.Y. Nie, Phys. Lett. B {\bf 776}, 115 (2018).

\bibitem{Nam}
C.H. Nam, Phys. Lett. B {\bf 797}, 134865 (2019); Gen. Rel. Grav. {\bf 51}, 104 (2019).

\bibitem{Parai}
D. Parai, D. Ghorai, and S. Gangopadhyay, Eur. Phys. J. C {\bf 80}, 232 (2020).

\bibitem{LiHF2011}
H.F. Li, R.G. Cai, and H.Q. Zhang, J. High Energy Phys. {\bf 04}, 028 (2011).

\bibitem{LuJW}
J.W. Lu, Y.B. Wu, T. Cai, H.M. Liu, Y.S. Ren, and M.L. Liu, Nucl. Phys. B {\bf 903}, 360 (2016).

\bibitem{LiuSC}
S.C. Liu, Q.Y. Pan, and J.L. Jing, Phys. Lett. B {\bf 765}, 91 (2017); arXiv:1610.02549 [hep-th].

\bibitem{Mohammadi}
M. Mohammadi and A. Sheykhi, Eur. Phys. J. C {\bf 79}, 743 (2019).

\bibitem{LaiCY}
C.Y. Lai, T.M. He, Q.Y. Pan, and J.L. Jing, Eur. Phys. J. C {\bf 80}, 247 (2020).

\bibitem{NieZY}
Z.Y. Nie and H. Zeng, J. High Energy Phys. {\bf 10}, 047 (2015).

\bibitem{Glavan}
D. Glavan and C.S. Lin, Phys. Rev. Lett. {\bf 124}, 081301 (2020); arXiv:1905.03601 [gr-qc].

\bibitem{HluPLB809}
H. Lu and Y. Pang, Phys. Lett. B {\bf 809}, 135717 (2020); arXiv:2003.11552 [gr-qc].

\bibitem{Hennigar}
R.A. Hennigar, D. Kubiznak, R.B. Mann, and C. Pollack, J. High Energy Phys. {\bf 07}, 027 (2020).

\bibitem{Fernandes}
P.G.S. Fernandes, P. Carrilho, T. Clifton, and D.J. Mulryne, Phys. Rev. D {\bf 102}, 024025 (2020).

\bibitem{Oikonomou}
V.K. Oikonomou and F.P. Fronimos, Class. Quantum Grav. {\bf 38}, 035013 (2021); arXiv:2006.05512 [gr-qc].

\bibitem{Aoki}
K. Aoki, M.A. Gorji, and S. Mukohyama, Phys. Lett. B {\bf 810}, 135843 (2020); arXiv:2005.03859 [gr-qc].

\bibitem{QiaoXY}
X.Y. Qiao, L. OuYang, D. Wang, Q.Y. Pan, and J.L. Jing, J. High Energy Phys. {\bf 12}, 192 (2020).

\bibitem{PanQY2012}
Q.Y. Pan, J.L. Jing, B. Wang, and S.B. Chen, J. High Energy Phys. {\bf 06}, 087 (2012).

\bibitem{Dong2014}
X. Dong, 
J. High Energy Phys.  {\bf 01}, 044 (2014); arXiv:1310.5713 [hep-th].

\bibitem{Wald1993}
R.M. Wald, Phys. Rev. D  {\bf 48}, 3427 (1993); arXiv:9307038 [gr-qc].

\bibitem{Wald1995}
V. Iyer and R.M. Wald, Phys. Rev. D {\bf 52}, 4430 (1995); arXiv:gr-qc/9503052.

\bibitem{Jacobson1994}
T. Jacobson, G. Kang, and R.C. Myers, Phys. Rev. D {\bf 49}, 6587 (1994); arXiv:gr-qc/9312023.


\end{thebibliography}
\end{document}